\documentclass[reprint, aps, pre, notitlepage, nofootinbib]{revtex4-2}

\usepackage{amsmath, amsfonts, amssymb, amsthm, mathtools}
\usepackage[utf8]{inputenc}
\usepackage[T1]{fontenc}
\usepackage{hyperref}
\usepackage{graphicx}
\usepackage[caption=false]{subfig}
\usepackage{lipsum}
\usepackage{xcolor}
\usepackage{empheq}
\usepackage[normalem]{ulem}
\theoremstyle{remark}

\theoremstyle{remark}

\theoremstyle{theorem}

\theoremstyle{theorem}

\newcommand{\LT}[1]{\widetilde{#1}}
\newcommand{\psm}{\psi_\mu}
\newcommand{\Psm}{\Psi_\mu}

\newcommand{\CV}{\mathrm{CV}}
\newcommand{\stkout}[1]{\ifmmode\text{\sout{\ensuremath{#1}}}\else\sout{#1}\fi}

\begin{document}
	
\title{Effects of mortality on stochastic search processes with resetting}
\author{Mattia Radice}
\email[Corresponding author: ]{mradice@pks.mpg.de}
\affiliation{Max Planck Institute for the Physics of Complex Systems, N\"{o}thnitzer Str. 38, 01187 Dresden, Germany}
\begin{abstract}
We study the first-passage time to the origin of a mortal Brownian particle, with mortality rate $ \mu $, diffusing in one dimension. The particle starts its motion from $ x>0 $ and it is subject to stochastic resetting with constant rate $ r $. We first unveil the relation between the probability of reaching the target and the mean first-passage time of the corresponding problem in absence of mortality, which allows us to deduce under which conditions the former can be increased by adjusting the restart rate. We then consider the first-passage time conditioned on the event that the particle reaches the target before dying, and provide exact expressions for the mean and the variance as functions of $ r $, corroborated by numerical simulations. By studying the impact of resetting for different mortality regimes, we also show that, if the average lifetime $ \tau_\mu=1/\mu $ is long enough with respect to the diffusive time scale $ \tau_D=x^2/(4D) $, there exist both a resetting rate $ r_\mu^* $ that maximizes the probability and a rate $ r_m $ that minimizes the mean first-passage time. However, the two never coincide for positive $ \mu $, making the optimization problem highly nontrivial.
\end{abstract}

\maketitle

\section{Introduction}
There is a plethora of situations in nature which can described by randomly moving objects in search for a fixed target. A short list of examples includes diffusion-controlled chemical reactions \cite{Kotomin1996}, diffusion in trapping environments \cite{Hug-I} and animals or even microorganisms in search for food \cite{Bell,Klafter1990}. These may be often formulated as \textit{first-passage problems} for a stochastic process, where one is mainly interested in the statistics of the \textit{first-passage time} (FPT), namely, the random variable that describes the time to reach the target for the first time \cite{Metzler2014}. A closely related quantity is the \textit{survival probability}, which can be defined as the probability that the target is not reached up to a certain time $ t $. 

Not surprisingly, being able to identify efficient strategies to minimize the time required to complete the search is of considerable importance, both practically and theoretically. In recent years, it has been pointed out that resetting intermittently a stochastic process to its initial condition can notably increase the search efficiency \cite{LubSinZuc-1993,MonZec-2002,EvaMaj-2011, EvaMaj-2014, KusMajSabSch-2014, ReuUrbKla-2014, KusGod-2015, RotReuUrb-2015, PalKunEva-2016, Reu-2016, PalReu-2017, CheSok-2018, EvaMaj-2018, Mas-2019, KusGod-2019, PalPra-2019, RayMonReu-2019,  RobHadUrb-2019, RayReu-2020, MerBoy-2021,  RAD-2021, RayReu-2021, SanDasNat-2021, SchBre-2021, SinSanIom-2021, TucGamMaj-2022, EliReu, RAD-2022, Ray-2022, JaiBoyPal-2022}. For instance, let us consider the paradigmatic example of a Brownian particle in one dimension, that starts its motion from $ x_0=x $ in search of a target located at the origin of coordinates. The first-passage time for this problem follows the Lévy-Smirnov distribution, with probability density function (PDF) given by \cite{Red}
\begin{equation}\label{key}
	f_0(x,t)=\frac{x}{\sqrt{4\pi Dt^3}}\exp\left(-\frac{x^2}{4Dt}\right),
\end{equation}
where $ D $ is the diffusion constant. One can check from this expression that the total probability $ \mathcal{E}(x) $ of reaching the target starting from $ x $ is one:
\begin{equation}\label{key}
	\mathcal{E}(x)=\int_{0}^{\infty}f_0(x,t)dt=1;
\end{equation}
nevertheless, if we consider the first moment of the distribution, namely, the \textit{mean first-passage time} (MFPT), due to the power-law decay $ t^{-3/2} $ of $ f_0(x,t) $, we obtain a divergent MFPT, independently of the initial distance from the origin. Hence, the diffusing particle reaches the target with probability one, but the expected time to perform this passage is infinite. This divergence is mostly due to the contribution of trajectories that initially diffuse away from the target, for which the first-passage time may be arbitrarily large. The probability of these contributions exhibits a relatively slow power-law decay, which, as already mentioned, is responsible for the divergence of the MFPT. Now it is intuitive that restarting the process from time to time, in such a way that the time scale of a diffusive excursion is cut off by the average time between two resetting events, will eliminate those ``bad'' contributions, and yield a finite MFPT.

The most studied type of resetting protocol is the one that assumes independent and identically distributed time intervals between resetting events, drawn from an exponential distribution:
\begin{equation}\label{eq:res_exp}
	\phi_r(t)=r\exp(-rt),
\end{equation}
where $ r $ is the (constant) resetting rate. A resetting protocol of this kind is also called \textit{Poissonian} resetting, due to the fact that the number of restarts up to time $ t $ follows Poissonian statistics. It has been shown that, introducing this restart protocol in the one-dimensional Brownian search problem, for any $ r>0 $ the system shows a finite MFPT, which is given by \cite{EvaMaj-2011}
\begin{equation}\label{eq:Intro_MFPT}
	\langle T_0(r;x)\rangle=\frac{\exp(x\sqrt{r/D})-1}{r}.
\end{equation}
Note that for both $ r\to0 $ and $ r\to\infty $ the previous expression yields diverging limits, which means that there exists an optimal value $ r_0^* $ of the rate where $ \langle T_0(r;x)\rangle $ reaches a minimum. The two divergences are easily explainable: For $ r\to\infty $, the average time between restarts tends to zero, hence most of the times the particle is reset before being able to reach the target, and the process is thus restarted again and again; for $ r\to0 $ instead, we must recover the behavior of the reset-free process, which indeed exhibits a diverging MFPT. We remark that also different kinds of resetting protocols have been considered, e.g., resetting at fixed time intervals \cite{BhaDeBRed-2016,PalReu-2017,BodSok-2020-BrResI,EliReu,BonPal-2021} or with time-dependent resetting rates \cite{PalKunEva-2016,EulMet-2016,RAD-2022}, which, however, do not change drastically the aforementioned features. For comprehensive reviews, see \cite{EvaMajSch-2020,PalKusReu-2020}.

A problem that has been partially overlooked in the literature concerns the impact stochastic resetting may have on the first-passage properties of \textit{evanescent} systems, e.g., an object, such as a radioactive species, subject to a natural decay process \cite{Zoi-2008}. Systems of this kind are ubiquitous in nature, especially in the context of chemical reactions or biochemical processes. Examples include phenomena such as luminescence quenching or excimer formation \cite{Tac-1980}, degradation mechanisms associated with morphogen gradients \cite{HorBerBar-2005,YusAbaLin-2010} and recombination of a pair of particles \cite{SanTac-1979}.

In practice, in these systems the diffusive objects can ``disappear'' after a certain amount of time, which in most cases is treated as a random variable drawn from a given PDF $ \psi(t) $. If the first moment of the distribution is well-defined, this introduces the constraint of a finite expected lifetime for the stochastic process.  Needless to say, this radically alters the first-passage properties of the system, as has been investigated largely in the literature: We refer here to studies on normal and anomalous diffusion \cite{AbaYusLin-2012,MeeRed-2015}, hopping models in one or more dimensions \cite{KolFis-2000,LohKru-2009,YusAbaLin-2013}, continuous-time random walks \cite{AbaYusLin-2010,AbaYusLin-2013}, space-time coupled random walks \cite{CamAbaMen-2015} and random walks on hierarchical graphs \cite{BalAbaAbi-2019}. Although in this paper we focus on first-passage problems, we point out that research on evanescence is not limited to them \cite{TucGamGup-2020}.

The first consequence of considering first-passage processes with finite expected lifetime is that the probability of reaching the target becomes smaller than one, as the particle may ``die'' before completing the search. Remarkably, introducing resetting can increase the probability that the search process will complete successfully, and one can even identify an optimal rate that maximizes the probability \cite{Bel-2018}. Another consequence is that the MFPT can never be infinite, because the time scale of the system is ruled by the average lifetime. Recall that introducing resetting at high rates for an \textit{immortal} system yields a diverging MFPT, which is a crucial observation to deduce the existence of a minimal MFPT. For evanescent systems, on the other hand, the MFPT cannot take arbitrarily large values, and thus it is expected that, even if resetting is introduced, it will remain bounded for each value of the resetting rate. Consequently, inferring the possibility of optimizing the MFPT becomes a much less trivial task, not least because one must keep in mind that the resetting mechanism also affects the probability of success.

In this paper we investigate the first-passage problem of a Brownian particle in one dimension that starts its motion from $ x>0 $, in search of a target located at the origin. The dynamics is subject to stochastic resetting at constant rate $ r $ and the particle can decay after a random time exponentially distributed, with mortality rate $ \mu $.  We first develop a general theory that allows us to compute the quantities of interest, such as the probability of a successful search or the conditional MFPT, for an arbitrary first-passage process subject to Poissonian resetting and an exponentially distributed decay time. In particular, In Sec. \ref{s:FPT} we derive the PDF of the first-passage time; in Sec. \ref{s:E} we obtain the probability of success and show that it can be expressed in terms of the MFPT of the corresponding problem in absence of mortality, which let us deduce the conditions under which resetting can increase the probability; in Sec. \ref{s:CMFPT} we compute the conditional MFPT and the variance, showing how both can be derived from the probability of success and highlighting the relation between the conditional MFPT and the moments of the FPT in absence of mortality. Then, in Sec. \ref{s:Line} we apply the previously developed theory to the aforementioned problem of one-dimensional search processes, showing that the probability of success and the conditional MFPT take on drastically different behaviors depending on the mortality rate considered. Finally, in Sec. \ref{s:Concl} we draw our conclusions.

\section{First-passage time distribution}\label{s:FPT}
As we have already mentioned, in this paper we will only consider exponential distributions for both the time intervals between resetting events and the ``decay'' time of the system. The PDF of the former is denoted with $ \phi_r(t) $ as in Eq. \eqref{eq:res_exp}, while the latter is drawn from
\begin{equation}\label{eq:psm_def}
	\psm(t)=\mu\exp(-\mu t),
\end{equation}
where $ \mu $ is the constant \textit{mortality rate}. The system is thereby characterized by the time scales $ \tau_r=1/r $ and $ \tau_\mu=1/\mu $, where $ \tau_r $ is the average time between resets and $ \tau_\mu $ is the average lifetime. We point out that that the restart mechanism here considered only affects the dynamics, and does not reset the decay time.
 
Let $ \Psm(t) $ be the probability that the system does not decay up to time $ t $, which, according to Eq. \eqref{eq:psm_def}, is given by
\begin{equation}\label{eq:Psi}
	\Psm(t)=\mu\int_{t}^{\infty}e^{-\mu \tau}d\tau=e^{-\mu t}.
\end{equation}
We consider an absorbing boundary located at the origin, and denote with $ Q(x,t;\mu,r) $ the survival probability up to time $ t $ of a particle with mortality $ \mu $, that starts its motion from $ x_0=x>0 $ and is reset at rate $ r $ to the starting position. We say that a mortal particle subject to stochastic resetting survives if it does not reach the target \emph{and} does not decay up to time $ t $, hence
\begin{equation}\label{key}
	Q(x,t;\mu,r)=q_r(x,t)\Psm(t),
\end{equation} 
where $ q_r(x,t) $ is the survival probability, for the same problem, of an immortal walker in presence of resetting.

Now, the difference between the particles that have survived up to time $ t $, and those who have survived up to time $ t+dt $, corresponds to the particles associated with the search processes that have stopped in the interval $ (t,t+dt) $, either because the target has been reached or because the particle has decayed. We thus define the PDF of the stopping time:
\begin{align}\label{key}
	F(x,t;\mu,r)&=-\frac{\partial Q(x,t;\mu,r)}{\partial t}\\
	&=f_r(x,t)\Psm(t)+q_r(x,t)\psm(t),\label{eq:F_time}
\end{align} 
where $ f_r(x,t) $ is the first-passage time of an immortal particle, defined as 
\begin{equation}\label{eq:f_rdef}
	f_r(x,t)=-\frac{\partial q_r(x,t)}{\partial t},
\end{equation}
and we used $ d\Psm(t)/dt=-\psm(t) $.

Clearly, the stopping time PDF is expressed as the sum of two contributions: The first term at the right-hand side (rhs) of Eq. \eqref{eq:F_time} corresponds to the PDF of the time of a successful search process, whereas the second one is the PDF of the stopping time due to the decay of the particle. Here we are interested only in the former:
\begin{equation}\label{key}
	F_1(x,t;\mu,r)=f_r(x,t)\Psm(t).
\end{equation}
If we consider its Laplace transform, by keeping in mind Eq. \eqref{eq:Psi}, we obtain
\begin{equation}\label{eq:F1_vs_fr}
	\widetilde{F}_1(x,s;\mu,r)=\int_{0}^{\infty}e^{-(s+\mu)t}f_r(x,t)dt=\widetilde{f}_r(x,s+\mu),
\end{equation}
where $ \widetilde{f}_r(x,p) $ is the Laplace transform of the first-passage time PDF for an immortal particle with resetting. It can be shown that, in the case of resetting performed at constant rate $ r $, $ \widetilde{f}_r(x,p) $ can be expressed as 
\cite{EvaMaj-2011,Reu-2016}:
\begin{equation}\label{eq:f_r_general}
	\widetilde{f}_r(x,p)=\frac{(p+r)\widetilde{f}_0(x,p+r)}{p+r\widetilde{f}_0(x,p+r)},
\end{equation}
where $ \widetilde{f}_0(x,p) $ the Laplace-transformed first-passage time PDF of the reset-free system, namely
\begin{equation}\label{key}
	\widetilde{f}_0(x,p)=\int_{0}^{\infty}e^{-pt}f_0(x,t)dt.
\end{equation}
It follows
\begin{equation}\label{eq:F1_exp}
	\widetilde{F}_1(x,s;\mu,r)=\frac{(s+\mu+r)\widetilde{f}_0(x,s+\mu+r)}{s+\mu+r\widetilde{f}_0(x,s+\mu+r)}.
\end{equation}

\section{Probability of reaching the target: Relation with the MFPT in absence of mortality}\label{s:E}
We now observe that
\begin{equation}\label{key}
	\widetilde{F}_1(x,0;\mu,r)=\int_{0}^{\infty}F_1(x,t;\mu,r)dt,
\end{equation}
is the total probability of hitting the target, which will henceforth be denoted by $ \mathcal{E}(\mu,r;x) $. In virtue of Eq. \eqref{eq:F1_vs_fr} and \eqref{eq:f_r_general}, this can be written as
\begin{equation}\label{eq:E_vs_f_r}
	\mathcal{E}(\mu,r;x)=\LT{f}_r(x,\mu)=\frac{(\mu+r)\widetilde{f}_0(x,\mu+r)}{\mu+r\widetilde{f}_0(x,\mu+r)}.
\end{equation}
Since it specifies the possibility of an outcome out of different alternatives (in our case, the particle can either reach the target or decay), $ \mathcal{E}(\mu,r;x) $  is a \textit{splitting probability} \cite{Red,KliVoiBen-2022}. Notably, Eq. \eqref{eq:E_vs_f_r} states that $ \LT{f}_r(x,\mu) $, which defines the distribution (in Laplace space, with Laplace variable $ \mu $) of the FPT for a system of immortal particles undergoing Poissonian resetting, is equal to the probability that the same system will reach the target once the constraint of a finite mean lifetime $ \tau_\mu=1/\mu $ is introduced. Hence, the moments of the FPT distribution in absence of mortality can be derived from $ \mathcal{E}(\mu,r;x) $, and in particular the MFPT can be obtained from
\begin{equation}\label{eq:MFPT_vs_E}
	\left.\frac{\partial\mathcal{E}(\mu,r;x)}{\partial\mu}\right|_{\mu=0}=\left.\frac{\partial\LT{f}_r(x,\mu)}{\partial\mu}\right|_{\mu=0}=-\langle T_0(r;x)\rangle,
\end{equation}
where $\langle T_0(r;x)\rangle$ is the MFPT. Indeed, from Eq. \eqref{eq:E_vs_f_r} we correctly obtain
\begin{equation}\label{eq:fpt_general}
	-\left.\frac{\partial\mathcal{E}(\mu,r;x)}{\partial\mu}\right|_{\mu=0}=\langle T_0(r;x)\rangle=\frac{1-\LT{f}_0(x,r)}{r\LT{f}_0(x,r)},
\end{equation}
as is known from the theory of first-passage with restarts, see for example \cite{Reu-2016}. By inverting the previous equation, we can write
\begin{equation}\label{key}
	\LT{f}_0(x,r)=\frac{1}{1+r\langle T_0(r;x)\rangle},
\end{equation}
and by plugging this expression in Eq. \eqref{eq:E_vs_f_r}, after the substitution $ r\to\mu+r $, we find a general relation between $ \mathcal{E}(\mu,r;x) $ and $ \langle T_0(r;x)\rangle $, which reads
\begin{equation}\label{eq:E_vs_MFPT}
	\mathcal{E}(\mu,r;x)=\frac{1}{1+\mu\langle T_0(\mu+r;x)\rangle}.
\end{equation}

Now note that for $ r=0 $
\begin{equation}\label{key}
	\mathcal{E}(\mu,0;x)=\widetilde{f}_0(x,\mu)=\frac{1}{1+\mu\langle T_0(\mu;x)\rangle},
\end{equation}
hence $ \mathcal{E}(\mu,0;x) $ is strictly smaller than one for $ \mu>0 $. Can we increase this probability by introducing resetting in the system? An immediate consequence of Eq. \eqref{eq:E_vs_MFPT} is that, if for some $ r>0 $ we have
\begin{equation}\label{key}
	\langle T_0(\mu+r;x)\rangle<\langle T_0(\mu;x)\rangle,
\end{equation}
then we surely increase the probability: $ \mathcal{E}(\mu,r;x)>\mathcal{E}(\mu,0;x) $. In particular, if $ \langle T_0(r;x)\rangle $ has a minimum at $ r^*>\mu $, then $ \mathcal{E}(\mu,r;x) $ has a maximum at $ r^*-\mu $. Indeed, by taking the derivative with respect to $ r $, one can write
\begin{equation}\label{key}
	\frac{\partial\ln\mathcal{E}(\mu,r;x)}{\partial r}=-\mu\frac{\partial\langle T_0(\mu+r;x)\rangle}{\partial r},
\end{equation}
hence the stationary points of $ \mathcal{E}(\mu,r;x) $ correspond to stationary points of $ \langle T_0(\mu+r;x)\rangle $. If the MFPT has the global minimum at $ r_0^* $, and no other stationary points, as it happens in the case of diffusion discussed in the Introduction, see Eq. \eqref{eq:Intro_MFPT}, then the splitting probability has the global maximum at $ r^*_\mu=r^*_0-\mu $. We therefore have the simple relation
\begin{equation}\label{eq:r_mu_vs_r}
	r_\mu^*=r_0^*-\mu,	
\end{equation}
which agrees to what obtained in \cite{Bel-2018} for the specific case of mortal Brownian searchers in one dimension.

The situation where the MFPT has a single critical point (a minimum) is commonly observed in many systems with resetting, for example the aforementioned normal diffusion \cite{EvaMaj-2011}, diffusion in logarithmic potentials \cite{RayReu-2020}, diffusion in an interval \cite{PalPra-2019}, nonlinear diffusion \cite{Che-2022} and also diffusion with resetting at time-dependent rates \cite{PalKunEva-2016,RAD-2022}. In such cases, Eq. \eqref{eq:r_mu_vs_r} tells us that $ r_\mu^* $ exists only if $ r_0^*>\mu $. It follows that $ \mathcal{E}(\mu,r;x) $ can be maximized only if the mortality rate is below a given threshold $ \mu^* $, which is precisely given by $ r_0^* $, i.e., $ \mu^*=r_0^* $. In terms of time scales, the average lifetime $ \tau_\mu $ must be larger than $ \tau_{0}^*=1/{r_0^*} $, which is the average time between resets for an immortal particle, when the process restarts at the optimal rate.

\section{Conditional first-passage time: Mean and variance}\label{s:CMFPT}
In this section we study the first two moments of the conditional FPT, which are obtained by averaging only over those processes that actually reach the target. We will find that both present significant differences from the corresponding moments in absence of mortality.

By definition, the first moment, namely, the conditional MFPT, is given by
\begin{equation}\label{key}
	\langle T_\mu(r;x)\rangle=\frac{\int_{0}^{\infty}tF_1(x,t;\mu,r)dt}{\int_{0}^{\infty}F_1(x,t;\mu,r)dt},	
\end{equation}
which can be conveniently computed from
\begin{equation}
	\langle T_\mu(r;x)\rangle=-\left.\frac{\partial\ln\widetilde{F}_1(x,s;\mu,r)}{\partial s}\right|_{s=0}.\label{eq:MFPT_vs_F1}
\end{equation}
By using Eq. \eqref{eq:F1_exp} we obtain an equation in which $ \langle T_\mu(r;x)\rangle $ is expressed in terms of $ \LT{f}_0(x,p) $ and its derivative:
\begin{multline}\label{eq:MFPT}
		\langle T_\mu(r;x)\rangle=\frac{1+r\LT{f}'_0(x,\mu+r)}{\mu+r\LT{f}_0(x,\mu+r)}-\frac{1}{\mu+r}\\
		-\frac{\partial \ln\LT{f}_0(x,\mu+r)}{\partial r},		
\end{multline}
where we used the notation
\begin{equation}\label{key}
	\LT{f}_0'(x,p)=\frac{\partial\LT{f}_0(x,p)}{\partial p}.
\end{equation}

Although the previous equation provides an exact expression in terms of the fundamental quantity $ \LT{f}_0(x,p) $, it is worth highlighting the dependence of the conditional MFPT on the moments of the FPT distribution in absence of mortality. From Eqs. \eqref{eq:F1_vs_fr} and \eqref{eq:E_vs_f_r}, we have $ \LT{F}_1(x,s;\mu,r)=\LT{f}_r(x,s+\mu) $ and $ \mathcal{E}(\mu,r;x)=\LT{f}_r(x,\mu) $. Hence Eq. \eqref{eq:MFPT_vs_F1} becomes
\begin{equation}\label{eq:cMFPT_vs_E}
	\langle T_\mu(r;x)\rangle=-\frac{\partial\ln\mathcal{E}(\mu,r;x)}{\partial\mu},	
\end{equation}
and by using Eq. \eqref{eq:E_vs_MFPT}, we obtain
\begin{equation}\label{eq:cMFPT_vs_MFPT}
		\langle T_\mu(r;x)\rangle=\frac{\langle T_0(\mu+r;x)\rangle+\mu\partial_\mu\langle T_0(\mu+r;x)\rangle}{1+\mu\langle T_0(\mu+r;x)\rangle}.
\end{equation}
Then, by using \cite{Reu-2016}
\begin{equation}\label{eq:Reu_der}
	2\frac{\partial\langle T_0(p;x)\rangle}{\partial p}=2\langle T_0(p;x)\rangle^2-\langle T_0^2(p;x)\rangle,
\end{equation}
we can write
\begin{equation}\label{eq:cMFPT_vs_Mom}
		\langle T_\mu(r;x)\rangle=\langle T_0(\mu+r;x)\rangle-\frac{\mu\langle T_0^2(\mu+r;x)\rangle}{2+2\mu\langle T_0(\mu+r;x)\rangle}.
\end{equation}

Let us first observe that for $ \mu=0 $ one correctly recovers $ \langle T_0(r;x)\rangle $. If instead we consider positive mortality rates $ \mu>0 $, for $ r=0 $ we obtain the conditional MFPT in absence of resetting, which reads
\begin{align}
	\langle T_\mu(0;x)\rangle&=\frac{\langle T_0(\mu;x)\rangle+\mu\partial_\mu\langle T_0(\mu;x)\rangle}{1+\mu\langle T_0(\mu;x)\rangle}\\
	&=-\frac{\partial\ln\LT{f}_0(x,\mu)}{\partial\mu},\label{eq:cMFPT_zero_r}	
\end{align}
where the second line can be derived by using the expression for $ \langle T_0(\mu;x)\rangle $ given by Eq. \eqref{eq:fpt_general}; alternatively, it follows from Eq. \eqref{eq:MFPT}. For $ r\to\infty $ instead, we use Eq. \eqref{eq:MFPT} and consider the three terms at the right-hand side separately. The first one tends to $ 1/\mu $: This can be shown by observing that $ \LT{f}'_0(x,p) $ is the Laplace transform of $ -tf_0(x,t) $, and from the initial value theorem we have
\begin{align}\label{key}
	\lim_{p\to\infty}p\LT{f}_0(x,p)&=\lim_{t\to0}f_0(x,t)\\
	-\lim_{p\to\infty}p\LT{f}'_0(x,p)&=\lim_{t\to0}tf_0(x,t).
\end{align}
The second one converges to zero, while Eq. \eqref{eq:cMFPT_zero_r} suggests that the third term corresponds to $ \langle T_{\mu+r}(0;x)\rangle $, hence for $ r\gg\mu $ we can approximate
\begin{equation}\label{eq:cMFPT_approx_limr}
	\langle T_\mu(r;x)\rangle\approx\tau_\mu+\langle T_r(0;x)\rangle.
\end{equation} 
For diffusion processes, we expect the last term at the rhs of this equation to vanish in the $ r\to\infty $ limit. Indeed, this represents the conditional MFPT of a diffusive particle with mortality rate $ \mu=r $ in absence of resetting. For very large $ r $, i.e., in the high-mortality limit, the main contribution to the average time to reach the target comes from very fast searchers \cite{MeeRed-2015}. It follows that the conditional MFPT converges to the average lifetime $ \tau_\mu=1/\mu $:
\begin{equation}\label{eq:lim_cMFPT}
	\lim_{r\to\infty}\langle T_\mu(r;x)\rangle=\tau_\mu.
\end{equation}

We point out that this result, valid for diffusion, is based on the fact that there is always a positive probability of covering arbitrarily large distances in arbitrarily short times. For processes subject to the constraint of a finite propagation speed $ c $, however, the term $ \langle T_r(0;x)\rangle $ is not expected to converge to zero. A counterexample is given by \emph{run-and-tumble} particles: The Laplace transform $ \LT{f}_0(p;x) $ for this case has been widely considered in the literature. In one dimension and with symmetric initial conditions, it reads \cite{EvaMaj-2018,MalJemKun-2018,Mas-2019,RAD-2021}
\begin{equation}\label{key}
	\LT{f}_0(x,p)=\frac1{2\gamma}\left[p+2\gamma-\sqrt{p(p+2\gamma)}\right]e^{-\tau_c\sqrt{p(p+2\gamma)}},
\end{equation}
 where $ \gamma $ is the \emph{tumbling rate} and $ \tau_c=x/c $. One can verify that Eq. \eqref{eq:cMFPT_zero_r} yields
 \begin{equation}\label{key}
 	\langle T_\mu(0;x)\rangle=\tau_c\left(\frac{1+\beta}{\sqrt{1+2\beta}}\right)+\frac{\tau_\mu}{2}\left(\frac{\sqrt{1+2\beta}-1}{1+2\beta}\right),
 \end{equation}
where $ \beta=\gamma/\mu $. In the limit of high mortality, one obtains
\begin{equation}\label{key}
	\lim_{\mu\to\infty}\langle T_\mu(0;x)\rangle=\tau_c,
\end{equation}
hence $ \langle T_\mu(0;x)\rangle $ does not vanish, and Eq. \eqref{eq:lim_cMFPT} in the case of run-and-tumble particles must be replaced by
\begin{equation}\label{key}
	\lim_{r\to\infty}\langle T_\mu(r;x)\rangle=\tau_\mu+\tau_c.
\end{equation}
 
In the following we will only focus on diffusion processes, for which Eq. \eqref{eq:lim_cMFPT} may be considered valid. The effects of finite propagation speeds will be investigated in forthcoming work. However, the important point is that, in contrast to the case of immortal searchers, both the limits $ \lim_{r\to0}\langle T_\mu(r;x)\rangle $ and $ \lim_{r\to\infty}\langle T_\mu(r;x)\rangle $ are finite. Then, deducing the existence of stationary points, e.g, a minimum, for the conditional MFPT becomes a highly nontrivial task. Of course, one can try to approach the problem analytically, but the expressions obtained for the derivative reveal a rather complex structure, which makes it difficult to deduce its general properties. In Appendix \ref{a:Der} we show that the critical points satisfy:
\begin{multline}\label{eq:cMFPT_st_points}
	\frac{1-\mu^2\partial_r\langle T_0(\mu+r;x)\rangle}{1+\mu\langle T_0(\mu+r;x)\rangle}\frac{\partial\langle T_0(\mu+r;x)\rangle}{\partial r}\\
	=-\mu\frac{\partial^2\langle T_0(\mu+r;x)\rangle}{\partial r^2},
\end{multline}
from which we deduce that an optimal rate that maximizes the probability of a successful search generally does not also minimize the conditional MFPT. Indeed, as we have seen in Sec. \ref{s:E}, the rate $ r_\mu^* $ that maximizes $ \mathcal{E}(\mu,r) $ exists if there is $ r_0^*>\mu $ that minimizes $ \langle T_0(r;x)\rangle $, in which case we have $ r_\mu^*=r_0^*-\mu $. If we plug this value in Eq. \eqref{eq:cMFPT_st_points}, the left-hand side (lhs) vanishes, but the rhs in general does not. Consequently, we expect that probability and mean first-passage time are optimized for distinct resetting rates.

We can extend the analysis to the second moment. In particular, observe that the variance
\begin{equation}\label{key}
	\sigma_{\mu}^2(r;x)=\langle T^2_\mu(r;x)\rangle-\langle T_\mu(r;x)\rangle^2,
\end{equation}
can be computed as
\begin{equation}\label{eq:sig_vs_F1}
	\sigma^2_\mu(r;x)=\left.\frac{\partial^2\ln\LT{F}_1(x,s;\mu,r)}{\partial s^2}\right|_{s=0},
\end{equation}
indeed
\begin{align}\label{key}
	\left.\frac{\partial^2\ln\LT{F}_1(x,s;\mu,r)}{\partial s^2}\right|_{s=0}&=\left.\frac{\partial^2\LT{F}_1(x,s;\mu,r)/\partial s^2}{\LT{F}_1(x,s;\mu,r)}\right|_{s=0}\nonumber\\
	&-\left[\left.\frac{\partial\LT{F}_1(x,s;\mu,r)/\partial s}{\LT{F}_1(x,s;\mu,r)}\right|_{s=0}\right]^2\nonumber\\
	&=\langle T^2_\mu(r;x)\rangle-\langle T_\mu(r;x)\rangle^2.
\end{align}
By using Eq. \eqref{eq:F1_exp} we find
\begin{multline}
	\sigma_\mu^2(r;x)=\left[\frac{1+r\LT{f}'_0(x,\mu+r)}{\mu+r\LT{f}_0(x,\mu+r)}\right]^2-\frac{1}{(\mu+r)^2}\\
	-\left[\frac{\partial \ln\LT{f}_0(x,\mu+r)}{\partial r}\right]^2+\frac{\mu\LT{f}_0''(x,\mu+r)/\LT{f}_0(x,\mu+r)}{\left[\mu+r\LT{f}_0(x,\mu+r)\right]},		
\end{multline}
and once again, by recalling the relation between $ \LT{F}_1(x,s;\mu,r) $ and $ \mathcal{E}(\mu,r;x) $, we can also derive the variance from the splitting probability. Indeed, Eq. \eqref{eq:sig_vs_F1} can be rewritten as
\begin{equation}
	\sigma^2_\mu(r;x)=\frac{\partial^2\ln\mathcal{E}(\mu,r;x)}{\partial\mu^2},
\end{equation}
and recalling Eq. \eqref{eq:cMFPT_vs_E}, we can write a simple relation between $ \sigma^2_\mu(r;x) $ and $ \langle T_\mu(r;x)\rangle $:
\begin{equation}\label{eq:cSig_vs_cMFPT}
	\sigma^2_\mu(r;x)=-\frac{\partial\langle T_\mu(r;x)\rangle}{\partial\mu}.
\end{equation}

For $ r=0 $, one has
\begin{align}\label{key}
	\sigma_\mu^2(0;x)&=-\frac{\partial\langle T_\mu(0;x)\rangle}{\partial\mu}\\
	&=\frac{\partial^2\ln\LT{f}_0(x,\mu)}{\partial\mu^2},
\end{align}
while in the limit $ r\to\infty $, we find that the variance can be estimated as
\begin{equation}\label{key}
	\sigma_{\mu}^2(r;x)\approx\frac1{\mu^2}+\sigma_{r}^2(0;x),
\end{equation}
hence, by using similar arguments as those used before, we may say that for diffusion processes the variance converges to $ \tau_\mu^2=1/\mu^2 $:
\begin{equation}\label{key}
	\lim_{r\to\infty}\sigma_\mu^2(r;x)=\tau_\mu^2,
\end{equation}
with the same caveat as Eq. \eqref{eq:lim_cMFPT}.
%which, together with Eq. \eqref{eq:lim_cMFPT}, suggests that in this limit the FPT follows an exponential distribution, with mean $ 1/\mu $ and variance $ 1/\mu^2 $.

In the following, it will be useful to evaluate the sign of the derivative $ \partial\langle T_\mu(r;x)\rangle/\partial r $ for $ r\to0 $, which indicates whether the introduction of resetting with infinitesimally small rate increases or decreases the conditional MFPT. In Appendix \ref{a:Der} we show that the condition of having a negative slope reads
\begin{equation}\label{eq:Cond_neg_slope}
	\CV^2_\mu>\left[1-\mathcal{E}(\mu,0;x)\right]\zeta_\mu^2-\mathcal{E}(\mu,0;x)\zeta_\mu,
\end{equation}
where we have introduced the coefficient of variation $ \CV_\mu$ of the reset-free mortal process
\begin{equation}\label{key}
	\CV_\mu=\frac{\sqrt{\langle T^2_\mu(0;x)\rangle-\langle T_\mu(0;x)\rangle^2}}{\langle T_\mu(0;x)\rangle}=\frac{\sigma_{\mu}(0;x)}{\langle T_\mu(0;x)\rangle},
\end{equation}
and the dimensionless quantity $ \zeta_\mu$
\begin{equation}\label{key}
	\zeta_\mu=\frac{\langle T_\mu(\infty;x)\rangle}{\langle T_\mu(0;x)\rangle}.
\end{equation}

\subsection{The gap}
As we have seen in the previous section, the conditional MFPT is finite for both $ r=0 $ and $ r\to\infty $. In general, there are no constraints on the relation between $ \langle T_\mu(0;x)\rangle $ and $ \langle T_\mu(\infty;x)\rangle =\tau_\mu$, and depending on the mortality rate, we can have $ \tau_\mu> \langle T_\mu(0;x)\rangle$ or vice versa. As a consequence, the existence of critical points can not be demonstrated with simple arguments, as it happens for immortal systems. It can be useful to introduce the gap $ \Delta_\mu(r)=\tau_\mu-\langle T_\mu(r;x)\rangle $, which can be expressed as
\begin{equation}\label{eq:Gap_r}
	\Delta_\mu(r)=\tau_\mu\frac{1-\mu^2\partial_\mu\langle T_0(\mu+r;x)\rangle}{1+\mu\langle T_0(\mu+r;x)\rangle},
\end{equation}
and is defined so that one has $ \tau_\mu>\langle T_\mu(r;x)\rangle $ when $ \Delta_\mu(r)>0 $. Note that the equation for the critical points, Eq. \eqref{eq:cMFPT_st_points}, can be rewritten
as
\begin{equation}\label{key}
	\Delta_\mu(r)\frac{\partial\langle T_0(\mu+r)\rangle}{\partial r}=-\frac{\partial^2\langle T_0(\mu+r)\rangle}{\partial r^2},
\end{equation}
which reveals how the position of these points depends on the gap and the shape of the MFPT in absence of resetting.

The quantity $ \Delta_\mu(r) $ carries information about the effects of resetting on the system. First, let us consider the \emph{initial gap}
\begin{equation}
	\Delta_\mu(0)=\tau_\mu\frac{1-\mu^2\partial_\mu\langle T_0(\mu;x)\rangle}{1+\mu\langle T_0(\mu;x)\rangle},\label{eq:Gap_0}
\end{equation}
which does not depend on $ r $ and is thus a property of the reset-free mortal system. A positive initial gap is observed when
\begin{equation}\label{key}
	\frac{\partial\langle T_0(\mu;x)\rangle}{\partial\mu}<\frac1{\mu^2}.
\end{equation}
Then, if we consider once again the typical situation of a single critical point for $ \langle T_0(\mu;x)\rangle $, the condition $ \Delta_\mu(0)>0 $ is necessary (but not sufficient) to optimize the probability of reaching the target with the introduction of resetting in the system. Indeed, we have seen in Sec. \ref{s:E} that in this case the probability can be increased only if $ \mu<r_0^* $, where $ r_0^* $ is the position of the minimum of $ \langle T_0(r;x)\rangle $. The derivative $ \partial\langle T_0(r;x)\rangle/\partial r $ evaluated at $ r=\mu $ must hence be negative, which implies a positive initial gap. Intuitively, we expect resetting to help if the particle can live long enough to complete the search before it dies, viz., if $ \tau_\mu>\langle T_\mu(0;x)\rangle $, which is equivalent to $ \Delta_\mu(0)>0 $. Otherwise, if $ \tau_\mu< \langle T_\mu(0;x)\rangle $, only particles that survive longer than average can complete the search, and it is intuitive that resetting in this condition makes the task harder, decreasing the probability of success.

It is thus meaningful to know the mortality rates for which there is a transition between a negative gap and a positive gap. These are given by the solutions of
\begin{equation}\label{key}
	\frac{\partial\langle T_0(\mu;x)\rangle}{\partial\mu}=\frac1{\mu^2}.
\end{equation}
In the typical situation, this equation has a solution, since the lhs is usually not bounded and increasing. Consequently, we expect to observe a negative initial gap for sufficiently high mortality rates.

Finally, the gap helps us deduce the existence of minima or maxima for $ \langle T_{\mu}(r;x)\rangle $. Indeed, let us assume that $ \Delta_\mu(0) $ is positive, so that the reset-free system has $ \tau_\mu>\langle T_\mu(0;x)\rangle $. We know that $\lim_{r\to\infty}\langle T_\mu(r;x)\rangle=\tau_\mu$, as shown in the previous section. Thus, if we can observe $ \Delta_\mu(r)<0 $, i.e., if the conditional MFPT is larger than the average lifetime for some $ r $, then there must exist $ r_M $ such that $ \langle T_\mu(r_M;x)\rangle $ is the global maximum:
\begin{equation}\label{key}
	\langle T_\mu(r_M;x)\rangle>\tau_\mu.
\end{equation}
Furthermore, if $ \Delta_\mu(0)>0 $, then a sufficient (but not necessary) condition for the existence of a minimum is that the derivative $ \partial\langle T_\mu(r;x)\rangle/\partial r $ be negative for $ r\to0 $. On the other hand, if $ \Delta_\mu(0) $ is negative, then $ \tau_\mu<\langle T_\mu(0;x)\rangle $, while we still have $\lim_{r\to\infty}\langle T_\mu(r;x)\rangle=\tau_\mu$. Then, in this case, a positive $ \partial_r\langle T_\mu(r;x)\rangle $ for $ r\to0 $ is a sufficient condition for the existence of a maximum, whereas a positive $ \Delta_\mu(r) $ for some $ r $ would imply the existence of a global minimum.

\section{First-passage outside the positive axis in one dimension}\label{s:Line}
In this section we will apply the theory developed earlier to the case of Brownian searchers in one dimension. We assume that the motion starts from $ x_0=x>0 $, while the target is located at the origin, and we call $ D $ the diffusion coefficient. The PDF of the FPT for this problem in Laplace space, without resetting and mortality, reads 
\begin{equation}\label{key}
	\widetilde{f}_0(x,p)=\exp\left(-x\sqrt{\frac pD}\right).
\end{equation}
When we introduce resetting and mortality, we can compute the probability of reaching the target from Eq. \eqref{eq:E_vs_f_r}, which yields
\begin{equation}\label{eq:E_line}
	\mathcal{E}(\mu,r;x)=\frac{\mu+r}{\mu e^{w}+r},
\end{equation}
where $ w=x\sqrt{(\mu+r)/D} $. For $ r=0 $ we recover the probability of hitting the target for mortal walkers in absence of resetting \cite{MeeRed-2015}, which we write as
\begin{equation}\label{key}
	\mathcal{E}(\mu,0;x)=\exp\left(-2\sqrt{\frac{\tau_D}{\tau_\mu}}\right),
\end{equation}
where we have introduced the diffusive time scale $ \tau_D=x^2/(4D) $. By analyzing the derivative with respect to $ r $ of Eq. \eqref{eq:E_line}, we find that $ \mathcal{E}(\mu,r;x) $ is increasing whenever
\begin{equation}\label{key}
	2-w>2e^{-w}.
\end{equation}
This condition may be satisfied only up to a value $ w_0^* $, which is of course the solution of $ 2-w=2e^{-w} $ given by
\begin{equation}\label{key}
	w_0^*=2+W_0\left(-\frac{2}{e^2}\right)\approx 1.5936.
\end{equation}
Here $ W_0(z) $ denotes the principal branch of the Lambert function. Therefore, the splitting probability is increasing for $ w<w_0^* $ and decreasing for $ w>w_0^* $, and so we have the global maximum at $ w=w_0^* $. The resetting rate maximizing $ \mathcal{E}(\mu,r;x) $ is thus
\begin{equation}\label{eq:ropt_Line}
	r_\mu^*=D\left(\frac{w_0^*}{x}\right)^2-\mu=r_0^*-\mu,
\end{equation}
as shown in Sec. \ref{s:E}. We recall that $ r_0^*=D(w_0^*/x)^2 $ is the optimal rate that minimizes the MFPT in absence of mortality, which is given by
\begin{equation}\label{eq:Line_MFPT}
	\langle T_0(r;x)\rangle=\frac{e^{x\sqrt{r/D}}-1}{r}.
\end{equation}
As we have already mentioned, $ r_\mu^* $ exists only if $ \mu $ is below the critical value $ r_0^* $. Otherwise, the derivative of $ \mathcal{E}(\mu,r;x) $ is negative for any $ r>0 $ and no critical point is seen. In terms of time scales, we need $ \tau_\mu>\alpha\tau_D $, with $ \alpha=(2/w_0^*)^2\approx1.575 $.

\begin{figure*}
	\subfloat{
		\includegraphics[width=.4\textwidth]{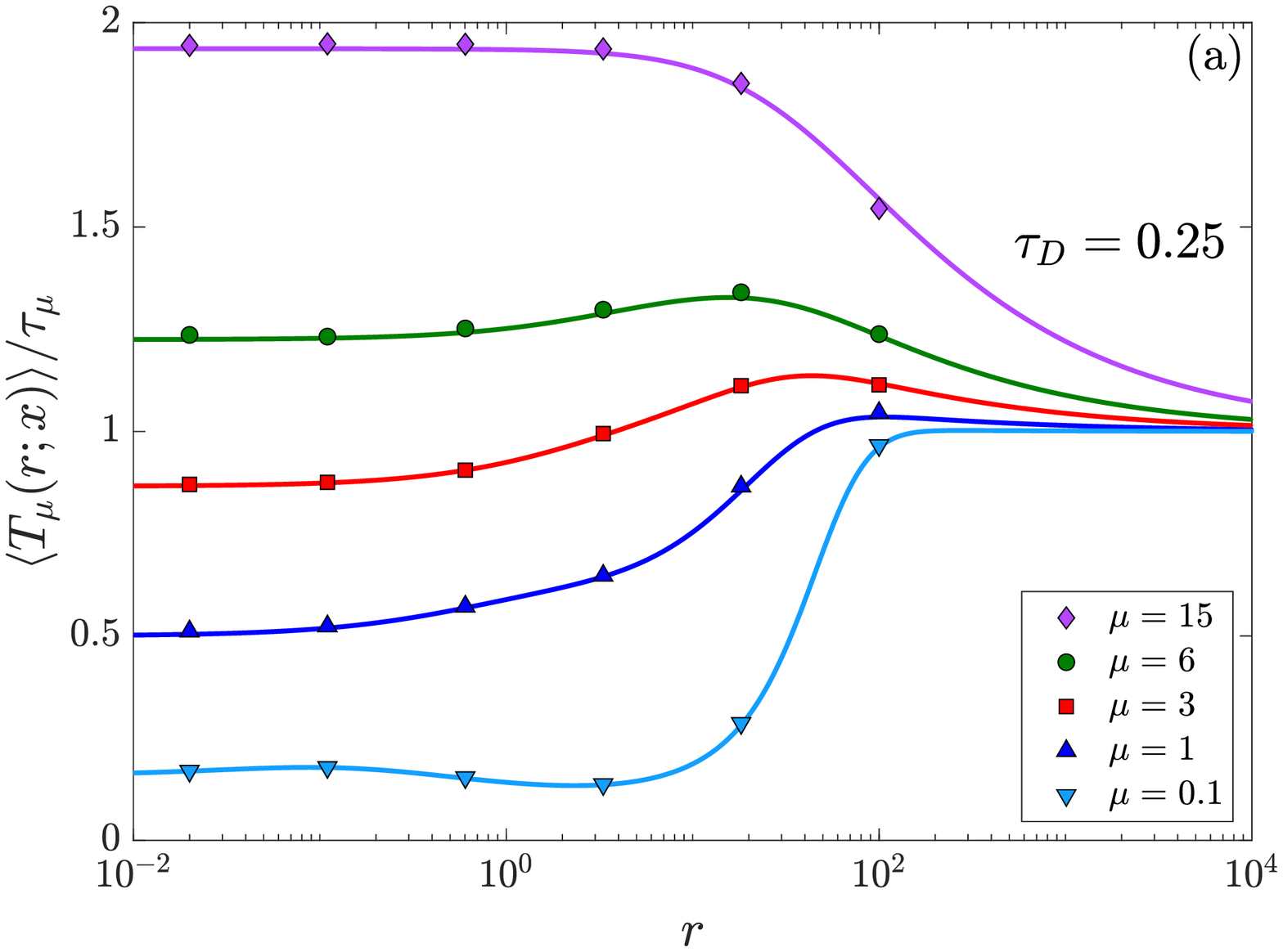}%
	}\quad
	\subfloat{
		\includegraphics[width=.4\textwidth]{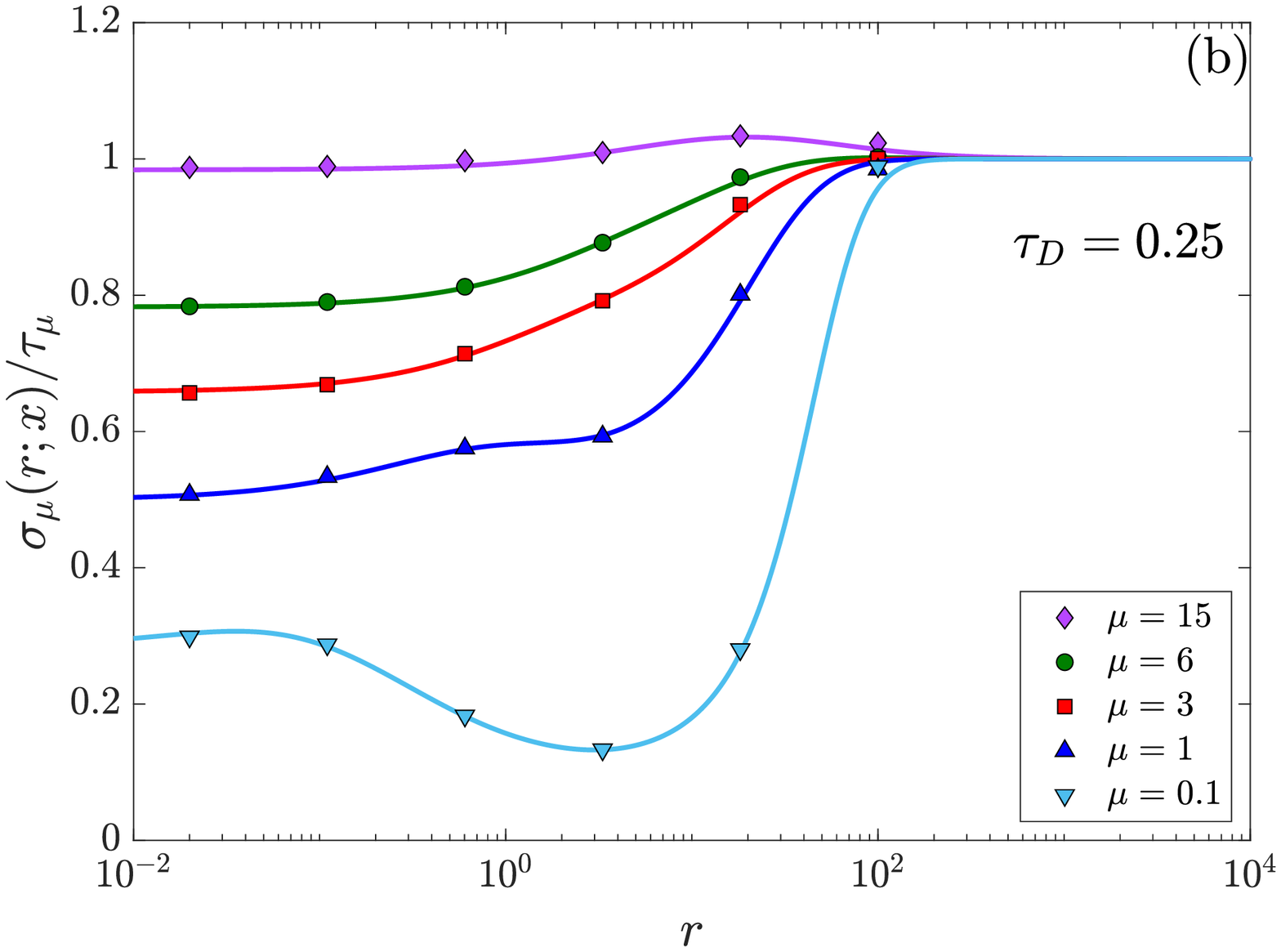}%
	}\hfill
	\caption{Numerical verification of the theoretical results for (a) the conditional MFPT, given by Eq. \eqref{eq:cMFPT_Line} and (b) the standard deviation, defined as the square root of Eq. \eqref{eq:Line_sig}: For each value of $ \mu $, the data sets are obtained by simulating $ N=2\cdot10^7 $ processes, with small time step $ \mathrm{d}t=5\cdot10^{-5} $, showing very good agreement. Note that for better readability, we plot $ \langle T_\mu(r;x)\rangle /\tau_\mu$ and $ \langle \sigma_\mu(r;x)\rangle /\tau_\mu$.}
	\label{fig:Line_Data}
\end{figure*}

We now consider the conditional MFPT, which can be obtained from the splitting probability by using Eq. \eqref{eq:cMFPT_vs_E}. A derivative with respect to $ \mu $ of $ \mathcal{E}(\mu,r;x) $ yields
\begin{equation}\label{eq:cMFPT_Line}
	\langle T_\mu(r;x)\rangle=\frac{r\left(e^w-1\right)+\mu we^w/2}{(\mu+r)(\mu e^w+r)},
\end{equation}
and a second derivative can be computed to obtain the variance:
\begin{multline}\label{eq:Line_sig}
	\sigma_\mu^2(r;x)=\frac{e^{2w}[(\mu+r)^2+(w-4)(\mu/2)^2]-r^2}{(\mu+r)^2(\mu e^w+r)^2}\\
	-\frac{re^w[rw+(w^2+3w+8)(\mu/4)]}{(\mu+r)^2(\mu e^w+r)^2}.
\end{multline}
The validity of these exact expressions is also confirmed by our numerical simulations, see Fig. \ref{fig:Line_Data}. Note that for $ r=0 $ we recover some previously known results \cite{MeeRed-2015}
\begin{align}
	\langle T_\mu(0;x)\rangle&=\frac{x}{2\sqrt{D\mu}}=\sqrt{\tau_D\tau_\mu}\label{eq:Line_T_r0}\\
	\sigma_\mu^2(0;x)&=\frac{x}{4\sqrt{D\mu^3}}=\frac{1}{2}\sqrt{\tau_D\tau_\mu^3},
\end{align}
while for $ r\to\infty $ we obtain the anticipated limits:
\begin{align}
	&\lim_{r\to\infty}\langle T_\mu(r;x)\rangle=\frac1\mu\label{eq:Line_Trinf}\\
	&\lim_{r\to\infty}\sigma_\mu^2(r;x)=\frac1{\mu^2}.
\end{align}

As we have already seen, we can get the behavior of the conditional MFPT curve from the gap $ \Delta_\mu(r) $. The sign of the gap is ruled by Eq. \eqref{eq:Gap_r}, and in particular we have a positive gap when
\begin{equation}\label{eq:Gap_der_condII}
	\frac{\partial\langle T_0(\mu+r;x)\rangle}{\partial\mu}<\frac{1}{\mu^2}.
\end{equation}
But from Eq. \eqref{eq:Line_MFPT} one can verify that for any fixed $ \mu $, a threshold can be found for the resetting rate $ r $ beyond which the derivative is always greater than $ 1/\mu^2 $. It follows that the gap is always negative beyond this threshold, and thus we have $ \langle T_\mu(r;x)\rangle>\tau_\mu $ for each $ r $ sufficiently large. If we assume that initially $ \Delta_\mu(0)>0 $, then there must exist a rate where $ \langle T_\mu(r;x)\rangle $ assumes its global maximum: In fact, the gap is initially positive, always negative after the threshold, and tends to zero for $ r\to\infty $. So it must have a global minimum, which implies the existence of a global maximum for $ \langle T_\mu(r;x)\rangle $. Note that, from Eqs. \eqref{eq:Line_T_r0} and \eqref{eq:Line_Trinf}, having $ \Delta_\mu(0)>0 $ corresponds to having
\begin{equation}\label{key}
	\mu<\frac{4D}{x^2},
\end{equation}
that is, $ \tau_\mu>\tau_D $. Thus, we can say that as long as the mortality is lower than $ 1/\tau_D $, the conditional MFPT curve always exhibits a global maximum. For higher mortality rates, on the other hand, we have $ \Delta_\mu(0)<0 $, then there is never any change in the sign of the gap, which remains negative for each $ r $. In this case, we can only conclude that $ \langle T_\mu(r;x)\rangle $ is bounded from below by $ \tau_\mu $: A maximum may still exist, but surely there is no global minimum.

Further information can be obtained by analyzing the condition given by Eq. \eqref{eq:Cond_neg_slope}, which says whether the introduction of an infinitesimal resetting rate decreases the conditional MFPT. The ratio $ \zeta_\mu $ now reads
\begin{equation}\label{key}
	\zeta_\mu=\frac{\langle T_\mu(\infty;x)\rangle}{\langle T_\mu(0;x)\rangle}=\frac{2}{x}\sqrt{\frac{D}{\mu}}=\sqrt{\frac{\tau_\mu}{\tau_D}},
\end{equation}
and $ \CV^2_\mu $ is equal to
\begin{equation}\label{key}
	\CV^2_\mu=\frac{\sigma_\mu^2(0;x)}{\langle T_\mu(0;x)\rangle^2}=\frac{1}{x}\sqrt{\frac{D}{\mu}}=\frac{1}{2}\zeta_\mu,
\end{equation}
hence Eq. \eqref{eq:Cond_neg_slope} reads
\begin{equation}\label{key}
	\left[1-\mathcal{E}(\mu,0;x)\right]\zeta_\mu^2-\left[\mathcal{E}(\mu,0;x)+\frac12\right]\zeta_\mu<0,
\end{equation}
or equivalently, since the values of $ \zeta_\mu $ are restricted to the positive real axis:
\begin{equation}\label{eq:Cond_xi}
	\zeta_\mu<\frac{1+2\mathcal{E}(\mu,0;x)}{2-2\mathcal{E}(\mu,0;x)}.
\end{equation}
Eq. \eqref{eq:Cond_xi} may be rewritten as
\begin{equation}\label{key}
	\zeta_\mu<\frac{e^{2/\zeta_\mu}+2}{2e^{2/\zeta_\mu}-2},
\end{equation}
which is solved for $ \zeta_\mu<\zeta^*\approx0.5371, $ yielding $ \mu>4D/(x\zeta^*)^2 $, that is, $ \tau_\mu<\beta\tau_D $, with $ \beta=(\zeta^*)^2\approx0.2884 $. Hence, as long as the mortality rate is below this value, the introduction of resetting with infinitesimally low $ r $ always hinders the conditional MFPT.

\begin{figure*}
	\subfloat{
		\includegraphics[width=.4\textwidth]{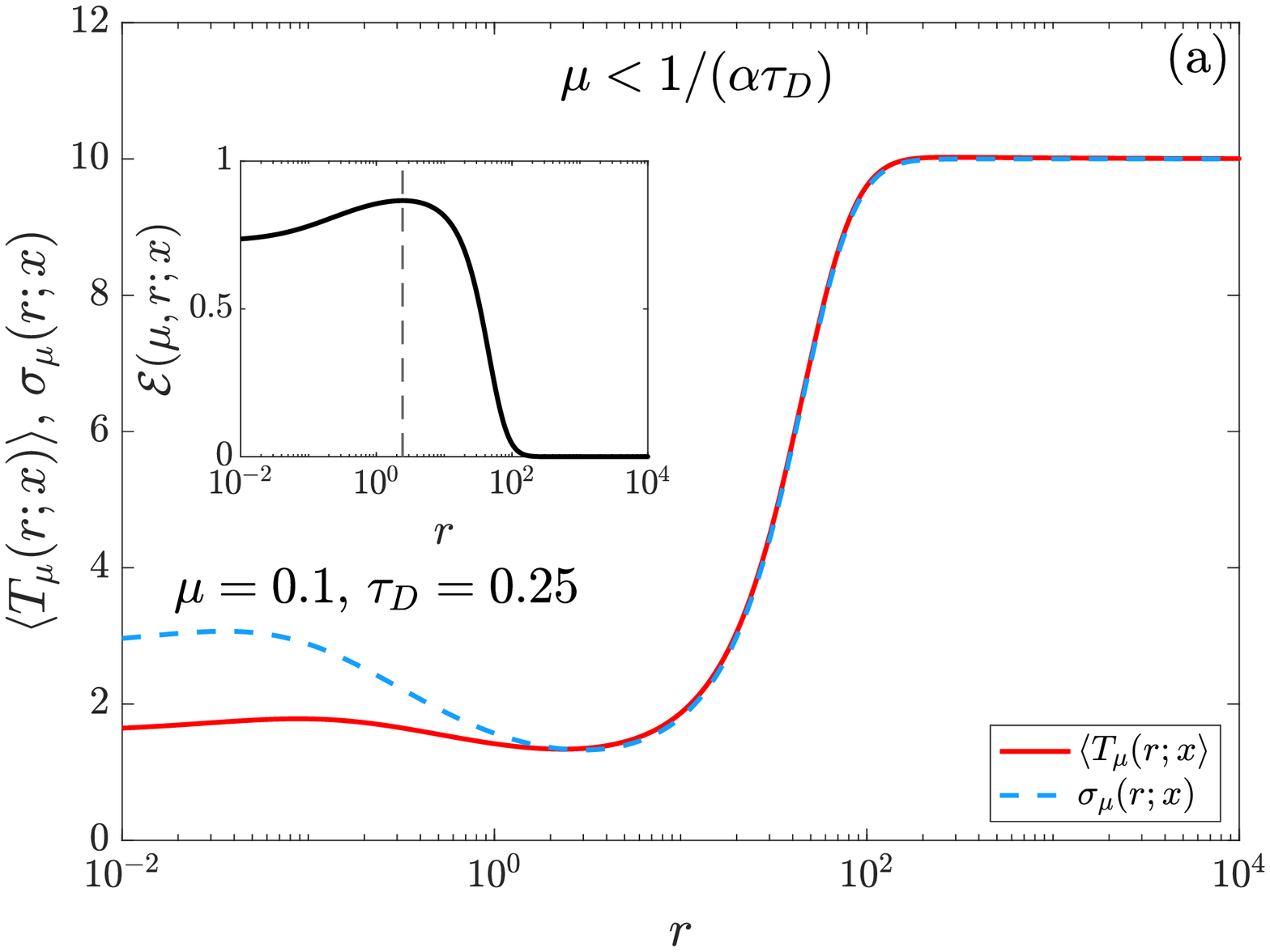}%
	}\quad
	\subfloat{
		\includegraphics[width=.4\textwidth]{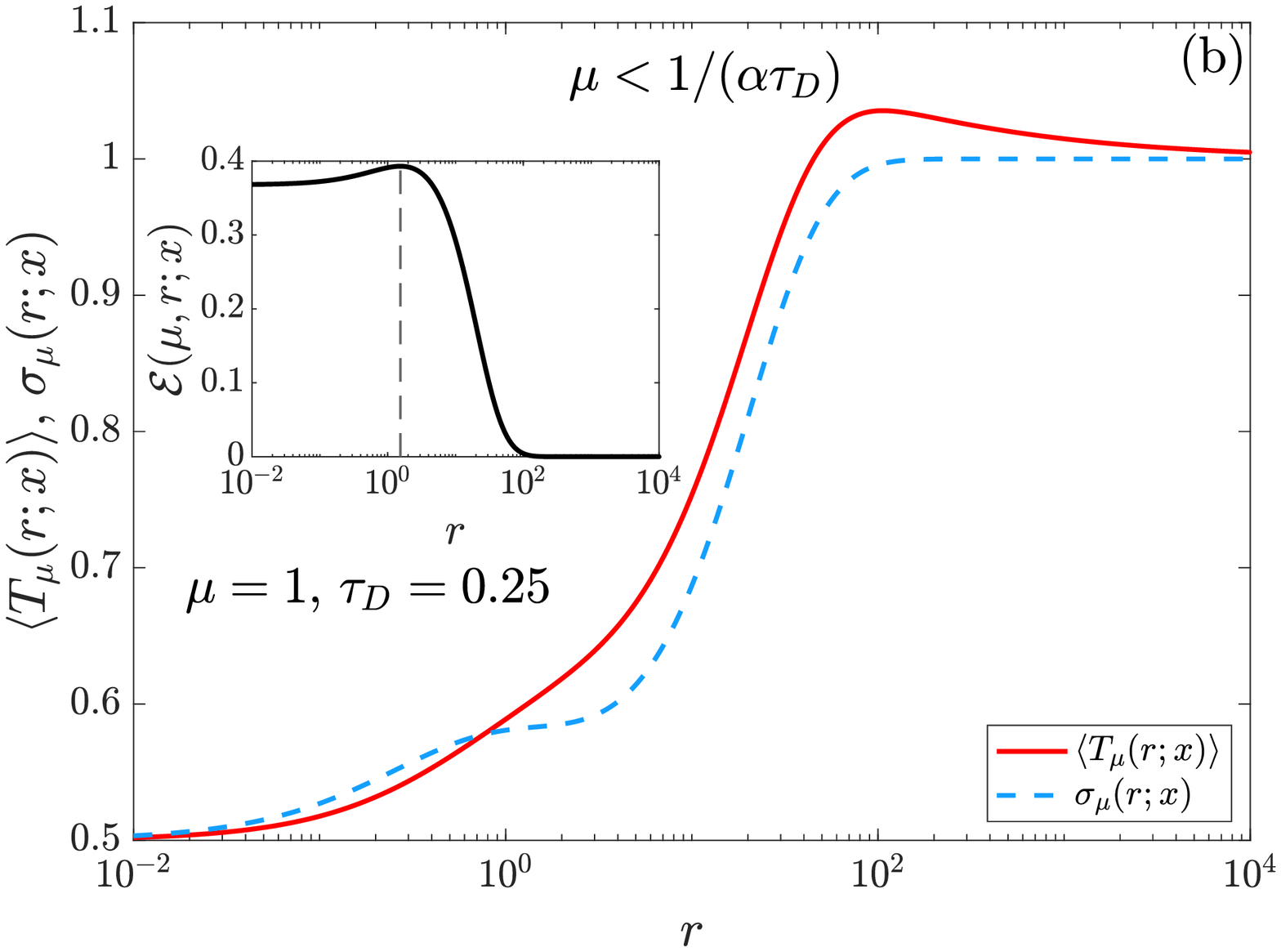}%
	}\hfill
	\subfloat{
		\includegraphics[width=.4\textwidth]{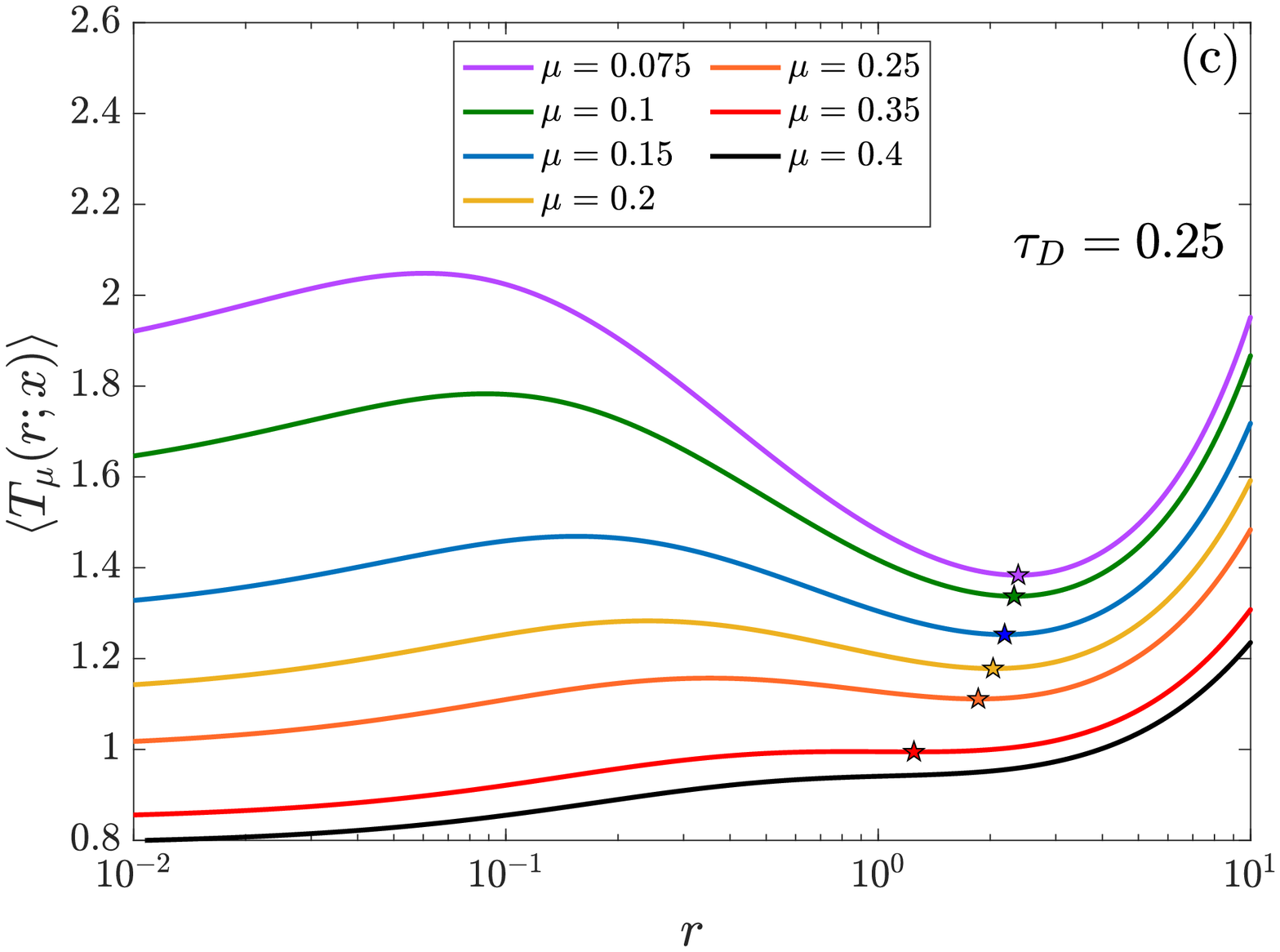}%
	}\quad
	\subfloat{
		\includegraphics[width=.4\textwidth]{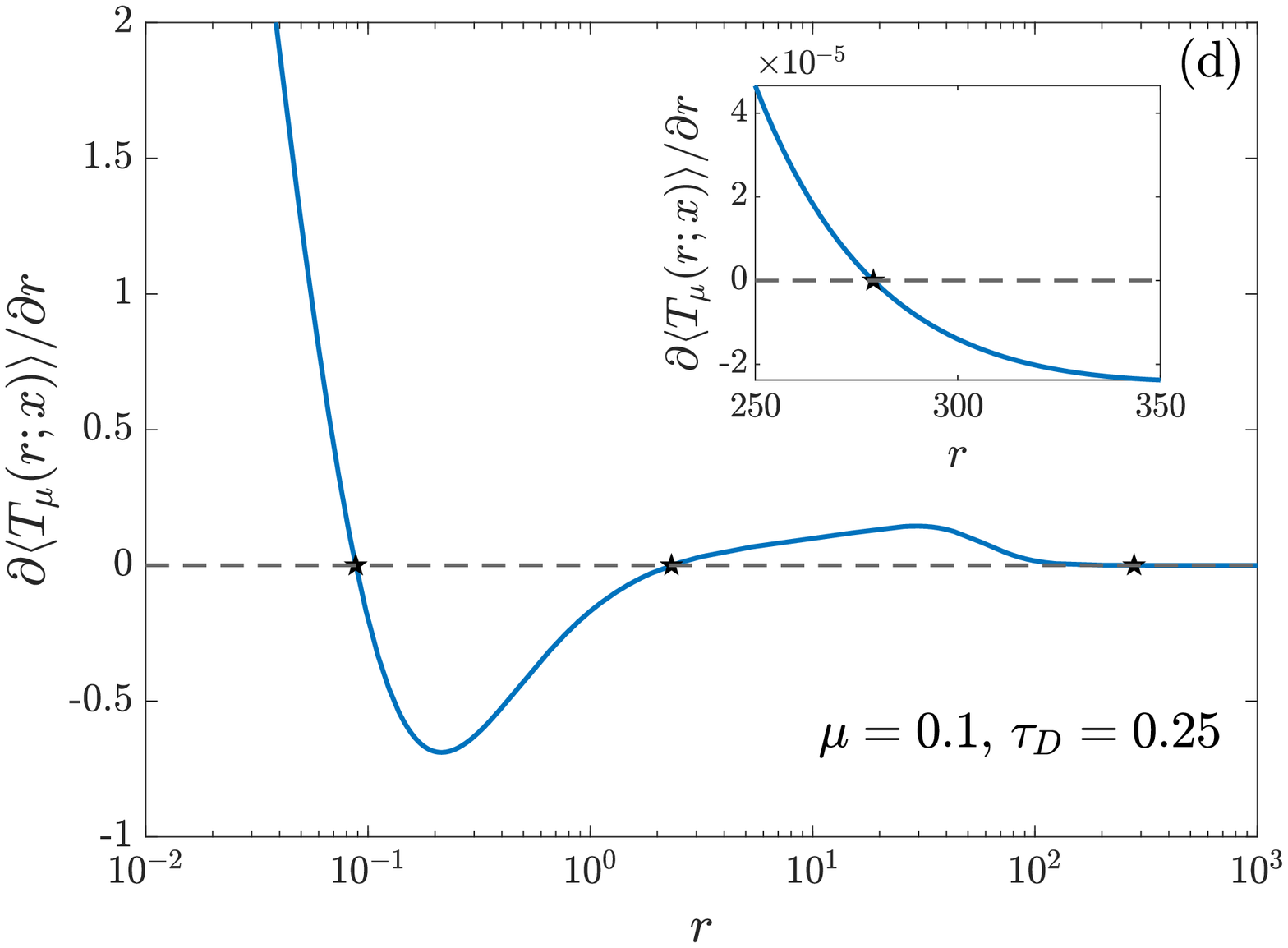}%
	}\hfill
	\caption{(a)-(b) Examples of conditional MFPT curves (solid red lines), given by Eq. \eqref{eq:cMFPT_Line}, and the corresponding standard deviations (dashed blue lines), i.e., the square root of Eq. \eqref{eq:Line_sig}, in the low mortality regime. The insets display the splitting probability, Eq. \eqref{eq:E_line}, with vertical dashed lines corresponding to the position of the optimal rate $ r_\mu^*=r_0^*-\mu $ that maximizes $ \mathcal{E}(\mu,r;x) $, see Eq. \eqref{eq:ropt_Line}. (c) Transition of the conditional MFPT curve in the low mortality regime. The markers represent the minimum of each curve, if present. For very low mortality rates, the curve has a global minimum for some positive $ r_m $. As $ \mu $ increases, the difference $ \langle T_\mu(0;x)\rangle-\langle T_\mu(r_m;x)\rangle $ becomes smaller and smaller, and eventually changes sign. Finally, for $ \mu $ large enough the minimum disappears. With our choice of the system parameters ($ x=D=1 $), the curves have a global minimum for $ \mu\lesssim0.15 $. The values of $ r_m $ are (curves from top to bottom): $ r_m\approx2.3774 $ for $ \mu=0.075 $, $ r_m\approx2.3172 $ for $ \mu=0.1 $ and $ r_m\approx2.1855 $ for $ \mu=0.15 $; for $ 0.15\lesssim\mu\lesssim0.35 $ the minima are just local, and located at $ r\approx2.0348 $ for $ \mu=0.2 $, $ r_m\approx1.8563 $ for $ \mu=0.25 $ and $ r_m\approx1.2474 $ for $ \mu=0.35 $. Note that in all cases we obtain $ r_m<r^*_\mu $. For $ \mu=0.4 $ and higher instead there is no minimum for positive values of $ r $. (d) Derivative with respect to $ r $ of the conditional MFPT, for the case $ \mu=0.1 $ illustrated in panel (a). The curve is computed from Eq. \eqref{eq:cMFPT_der_vs_Sig} and equal to zero for three values of $ r $, which are: $ \overline{r}_M\approx0.0881 $, $ r_m\approx2.3172 $ and $ r_M\approx279.1657 $, corresponding to three critical points of $ \langle T_\mu(r;x)\rangle $.  The inset displays more clearly the presence of the third zero.}
	\label{fig:Line_cMFPT1}
\end{figure*}

\begin{figure*}
	\subfloat{
		\includegraphics[width=.4\textwidth]{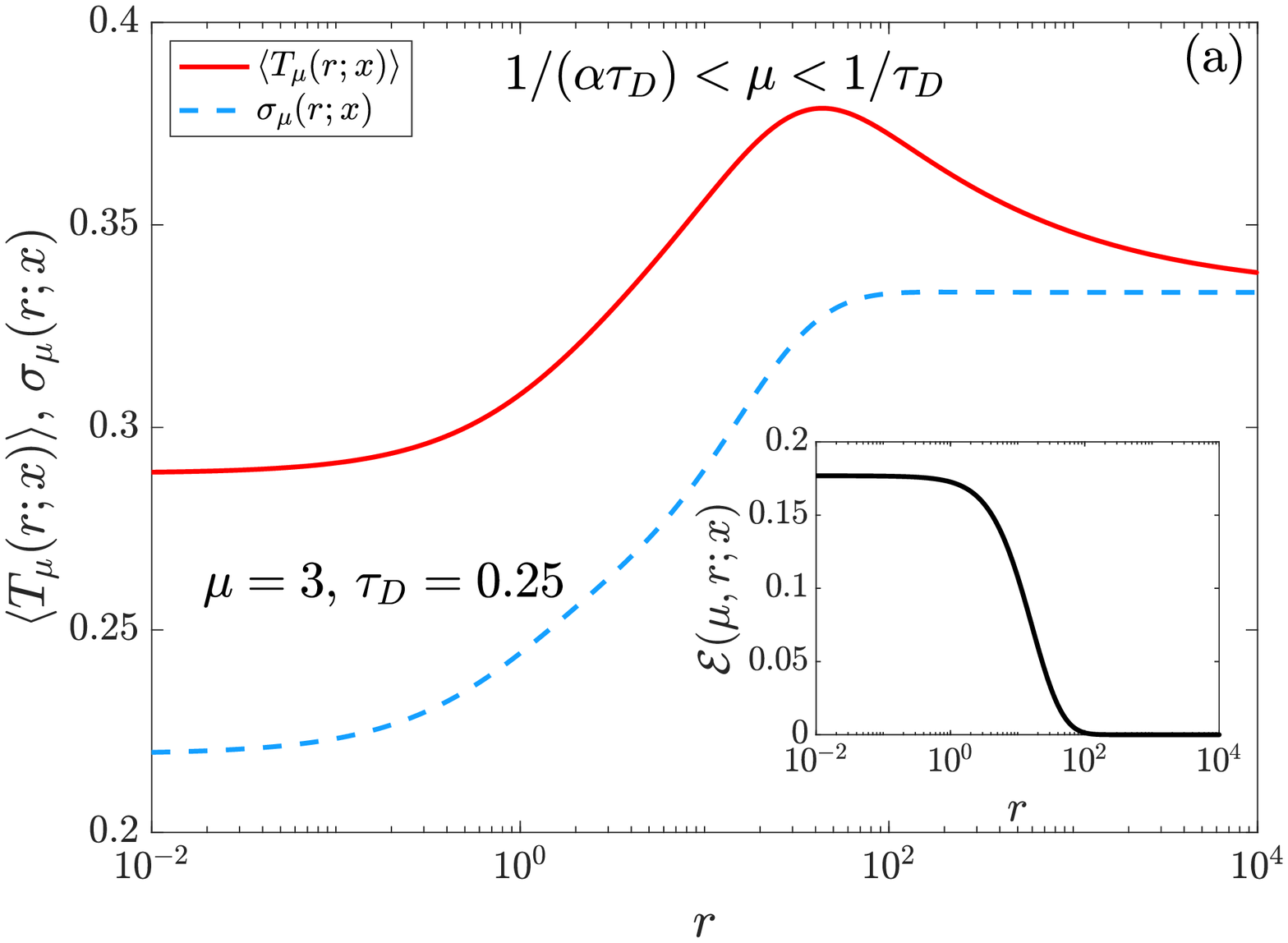}%
	}\quad
	\subfloat{
		\includegraphics[width=.4\textwidth]{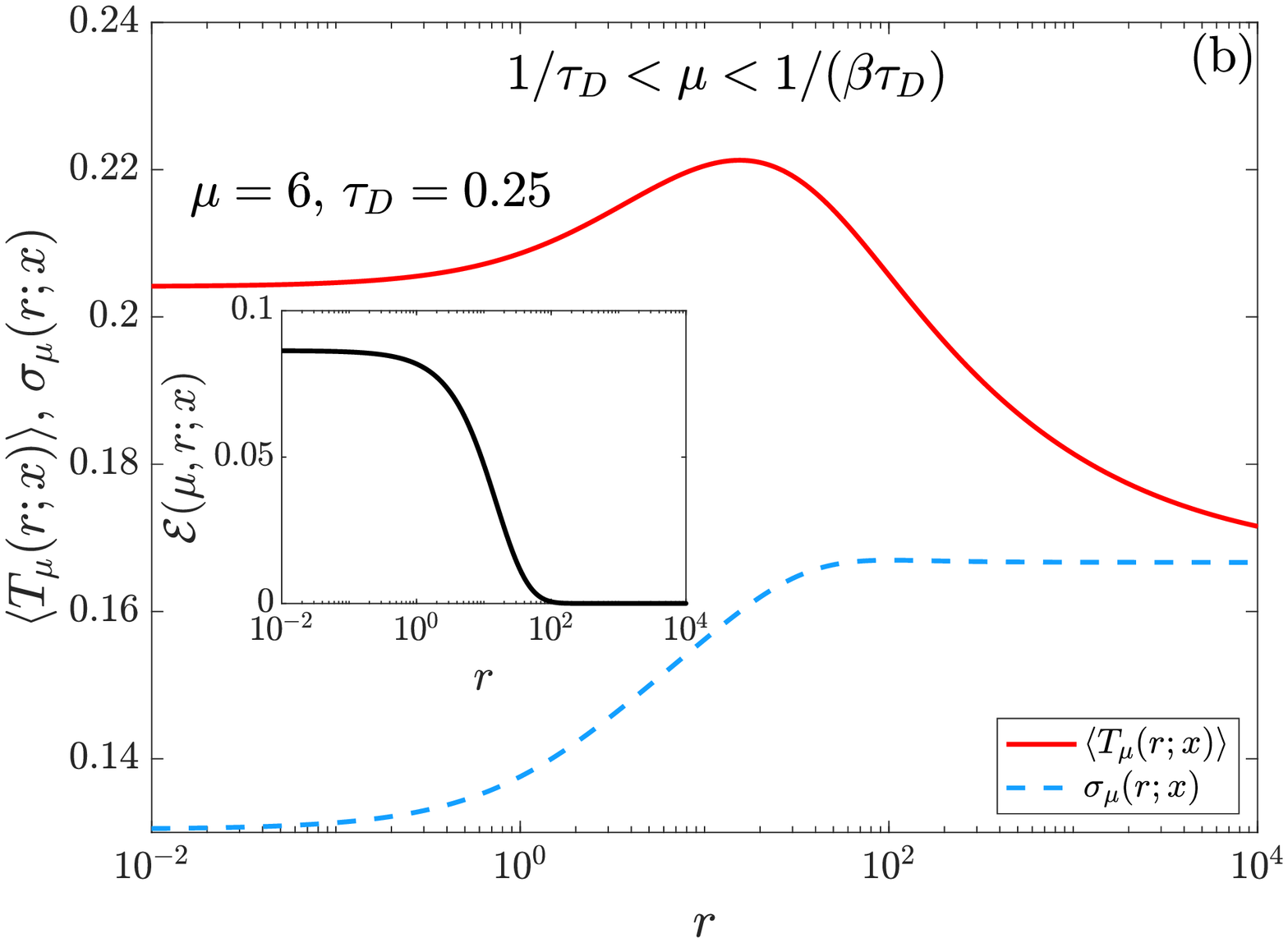}%
	}\hfill
	\subfloat{
		\includegraphics[width=.4\textwidth]{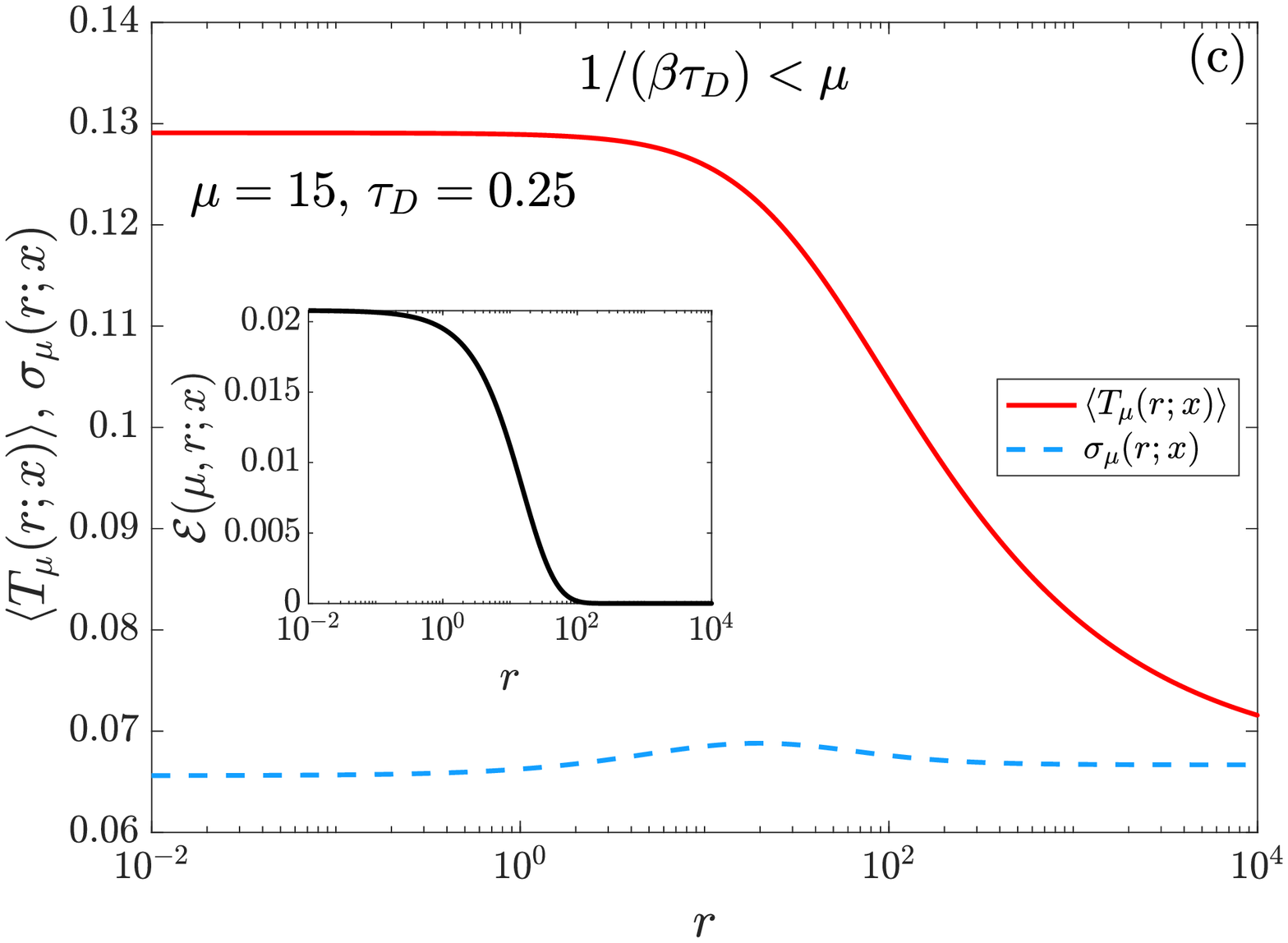}%
	}\quad
	\subfloat{
		\includegraphics[width=.4\textwidth]{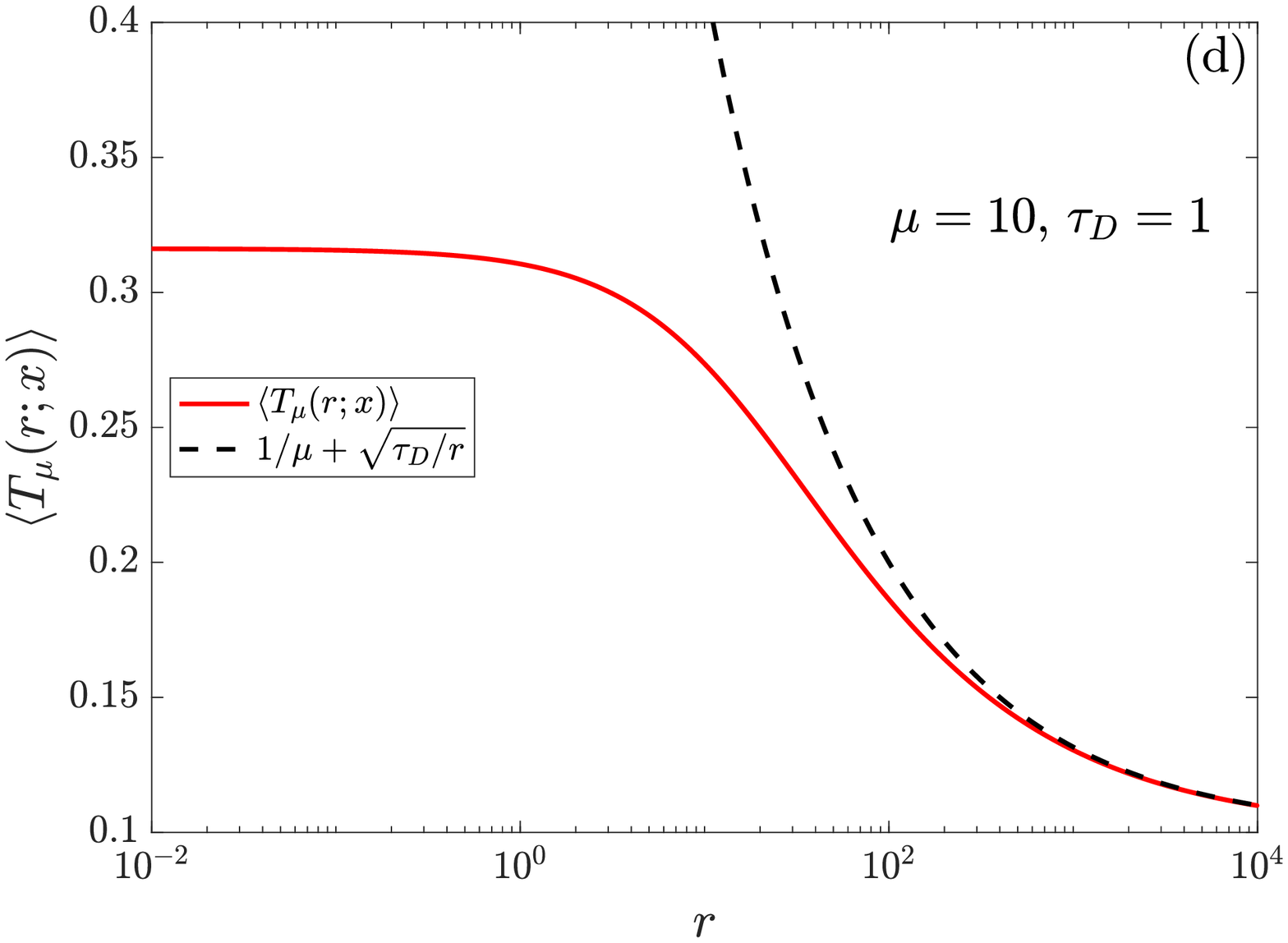}%
	}\hfill
	\caption{Examples of conditional MFPT curves (solid red lines), given by Eq. \eqref{eq:cMFPT_Line}, and the corresponding standard deviations (dashed blue lines), i.e., the square root of Eq. \eqref{eq:Line_sig}, in the (a) intermediate, (b) high and (c) extreme mortality regime. The insets display in each case the splitting probability, Eq. \eqref{eq:E_line}. (d) Comparison between the conditional MFPT (solid red curve) and its large-$ r $ asymptotic approximation (dashed black line) given by Eq. \eqref{eq:Asympt_r}, in the extreme mortality regime (with $ \tau_\mu=0.1 $ and $ \tau_D=1 $). The approximation becomes valid for $ r\gtrsim10^2 $, i.e., when $ \tau_r=1/r $ is much smaller than $ \tau_\mu $.}
	\label{fig:Line_cMFPT2}
\end{figure*}

The previous analysis thus revealed the existence of different mortality regimes that modify the behavior of $ \langle T_\mu(r;x)\rangle $, which are summarized in Figs. \ref{fig:Line_cMFPT1}-\ref{fig:Line_cMFPT2}:
\begin{enumerate}
	\item[i)] For $ \mu<1/(\alpha\tau_D) $, with $ \alpha\approx1.575 $, the system is in the \emph{low mortality} regime [Figs. \ref{fig:Line_cMFPT1}(a)-(b)]. The average lifetime is sufficiently large, i.e., $ \tau_\mu>\alpha\tau_D $, and the probability of reaching the target can thus be maximized by introducing resetting at the optimal rate $ r_\mu^*=1/(\alpha\tau_D)-\mu $. The initial gap $ \Delta_\mu(0) $ is positive, therefore $ \langle T_\mu(r;x)\rangle $ is initially smaller than the average lifetime; however, it becomes always larger than $ \tau_\mu $ for $ r $ large enough and attains its maximum for some positive $ r_M $. For resetting rates higher than $ r_M $, the curve decreases monotonically toward $ \tau_\mu $. The conditional MFPT increases when we consider resetting at infinitesimally small rates, hence the existence of minima is not guaranteed, see for instance Fig. \ref{fig:Line_cMFPT1}(b), where we have chosen $ \mu=1 $ and no minimum is seen. However, the curve can actually assume the global minimum for some positive $ r_m $, if we consider $ \mu $ small enough, as shown in Fig. \ref{fig:Line_cMFPT1}(a), with $ \mu=0.1 $. To better investigate this transition, in Fig. \ref{fig:Line_cMFPT1}(c) we consider different values of $ \mu\ll1/(\alpha\tau_D) $ and evaluate numerically the existence of a minimum. In the caption, we report the $ r_m $ that we obtain for the chosen values of $ \mu $. One can verify that in every case $r_m< r^*_\mu $. With our choice of the system parameters ($ x=D=1 $), for mortality rates $ \mu\lesssim0.15 $ there is in effect a global minimum. For higher rates, the minimum is just local, standing at higher values than $ \langle T_\mu(0;x)\rangle $. For $ \mu\gtrsim0.35 $ the minimum disappears. Note that, since the curve is increasing for very low resetting rates, the minimum, if present, is always preceded by a local maximum, see for example Fig. \ref{fig:Line_cMFPT1}(d), where we plot the derivative of the curve for $ \mu=0.1 $. Finally, Fig. \ref{fig:Line_cMFPT1}(a) seems to suggest that the position of the minimum coincides with the rate $ r_c $ at which the conditional MFPT and the standard deviation are equal, i.e., $ \sigma_{\mu}(r_c;x)=\langle T_\mu(r_c;x)\rangle $. However, closer analysis reveals that it is not the case: For $ \mu=0.1 $, we find numerically $ r_m\approx2.3172 $ and $ r_c\approx2.3884 $. We discuss this in the next section.
	\item[ii)] For $ 1/(\alpha\tau_D)<\mu<1/\tau_D $, the system is in the \emph{intermediate mortality} regime. The average lifetime is larger than the diffusive time scale, but not large enough to make resetting effective to increase the probability of reaching the target, i.e., $ \tau_D<\tau_\mu<\alpha\tau_D $. The typical behavior of the conditional MFPT is presented in Fig. \ref{fig:Line_cMFPT2}(a). The initial gap $ \Delta_\mu(0) $, as in the previous regime, is positive, and indeed the curve reaches a maximum for some $ r_M $, above which it decreases monotonically and finally converges to $ \tau_\mu $.
	\item[iii)] For $ 1/\tau_D<\mu<1/(\beta\tau_D) $, with $ \beta\approx 0.2884 $, the system is in the \emph{high mortality} regime. The average lifetime is now shorter than the diffusive time scale, but still larger than $ \beta\tau_D $, i.e., $ \beta\tau_D<\tau_\mu<\tau_D $. An example of MFPT curve is displayed in Fig. \ref{fig:Line_cMFPT2}(b). The initial gap is now negative, that is, $ \langle T_\mu(0;x)\rangle>\tau_\mu$, meaning that in the system without resetting only a relatively small set of particles with a longer than average lifetime can complete the search. By introducing resetting into the system, we gradually eliminate from this set the particles that decay the fastest, decreasing the probability of success but increasing $ \langle T_\mu(r;x)\rangle $. Hence, for low values of $ r $, the effect is similar to the previous cases: The curve initially increases with $ r $, reaches its maximum for some $ r_M $, and for $ r>r_M $ it always decreases towards $ \tau_\mu $.
	\item[iv)] For $ \mu>1/(\beta\tau_D) $, we have the \emph{extreme mortality} regime. The average lifetime is so short, i.e., $ \tau_\mu<\beta\tau_D $, that for any value of $ r $ the system is essentially in the same situation we observed for the previous cases when $ r>r_M $, see Fig. \ref{fig:Line_cMFPT2}(c). The initial gap is negative and the introduction of infinitesimally small resetting rates reduces the conditional MFPT. As a consequence, we observe a curve that decreases monotonically toward $ \tau_\mu $. This happens because the particles decay so fast that resetting, even for infinitesimally small resetting rates, progressively ``filters out'' the slower search processes.
\end{enumerate}

Now that we have a fairly complete picture, we can try to understand qualitatively how resetting affects the first-passage properties of the system. It should be clear that the behavior of the conditional MFPT as a function of $ r $ varies with the duration of the average lifetime $\tau_\mu$ with respect to the diffusive time scale $\tau_D$.

Let us consider first the low mortality regime depicted in Fig. \ref{fig:Line_cMFPT1}(a), where we have $ \tau_\mu\gg\alpha\tau_D $. For $ r=0 $, the splitting probability is close to one, see Eq. \eqref{eq:E_line}, hence the few particles that do not reach the target are those that decay during long diffusive excursions away from it. When we introduce resetting at very low rates, we occasionally ``save'' one of these particles, returning it to the starting point and allowing it to hit the target before decaying. We hence increase the probability of a successful search. However, since the rate is low, we let a particle diffuse for a long time before resetting it. Therefore, this particle brings a contribution that increases the conditional MFPT. If we consider higher resetting rates, we save more and more particles, but we increase $ \langle T_\mu(r;x)\rangle $ even more, up to the point where we reach a local maximum. With even higher values of $ r $, we reach the domain of rates where resetting also decreases $ \langle T_\mu(r;x)\rangle $. This basically happens because $\tau_r$ is such that the particles diffusing away from the target are not allowed to move too far away, but at the same time, those that diffuse toward the target are not hindered by the resetting mechanism. We then arrive at a rate $ r_m $ that minimizes the conditional MFPT. Slightly above $ r_m $, we are still saving more and more particles (the splitting probability is still increasing), but we are also starting to increase $ \langle T_\mu(r;x)\rangle $. When we reach $ r_\mu^* $, we obtain the largest possible $ \mathcal{E}(\mu,r;x) $. For $ r>r^*_\mu $ we are gradually decreasing the splitting probability, basically because we are considering such high rates that some of the particles that would reach the target without resetting are no longer able to do so. Indeed, eventually $ \mathcal{E}(\mu,r;x)<\mathcal{E}(\mu,0;x) $. At the same time, we are also increasing $ \langle T_\mu(r;x)\rangle $, because we are resetting the particles more and more. By considering higher and higher rates, we eventually reach a regime where $ \langle T_\mu(r;x)\rangle>\tau_\mu $. This means that now, due to the resetting mechanism, only particles that survive longer than average can complete the search. However, since the system cannot sustain arbitrarily long lifetimes, there must be a maximum value that we can observe for $ \langle T_\mu(r;x)\rangle $, which is obtained for some $ r_M $. Above $ r_M $, the conditional MFPT decreases monotonically toward $ \tau_\mu $: We can understand this behavior by considering that if we take a particle with $ \tau_\mu\gg\tau_r $, where $ \tau_r $ is the average time between two resetting events, then it is reset approximately $ N_r\approx\tau_\mu/\tau_r $ times before decaying. We can then interpret the search process as a sequence of $ N_r $ Bernoulli trials, where each trial is the search process of a particle with mortality rate $ \mu=r $ \emph{without resetting}, and we imagine to stop the trials at the first success. The probability of success of each single trial is of course $ q=\mathcal{E}(r,0;x)=\exp(-2\sqrt{\tau_D/\tau_r}) $, and the expected number of trials before the first success is $ \overline{N_r^s}=1/q=\exp(2\sqrt{\tau_D/\tau_r}) $. Hence, an exponentially large number of trials is expected before observing the first success. But the number of trials can not be arbitrarily large, because $ \tau_\mu $ is finite, and thus in most cases the process is concluded unsuccessfully. Therefore, even for those few processes that end in success, a high number of attempts is expected, and the target is most likely reached after about $ N_r $ failed attempts. This means that in this very high resetting regime, $ \langle T_\mu(r;x)\rangle $ can be estimated as
\begin{align}
	\langle T_\mu(r;x)\rangle &\approx\tau_\mu+\langle T_r(0;x)\rangle\\
	&\approx\tau_\mu+\sqrt{\tau_D\tau_r}.\label{eq:Asympt_r}
\end{align}
Indeed, one can verify that the same estimate is obtained from an asymptotic expansion of Eq. \eqref{eq:cMFPT_Line} for large $ r $. This is also illustrated in Fig. \ref{fig:Line_cMFPT2}(d), where we plot $ \langle T_\mu(r;x)\rangle $ for $ \tau_\mu=0.1 $ and $ \tau_D=1 $. Consequently, $ \tau_\mu<\beta\tau_D $ and we are thus in the extreme mortality regime. Note that for low $ r $, resetting has little effect on $ \langle T_\mu(r;x)\rangle $, because $ \tau_\mu\gg\tau_r $ and so the particle is unlikely to undergo resetting before decaying. For high rates instead, the approximation becomes valid and we observe a power-law decay of $ \langle T_\mu(r;x)\rangle \sim \tau_\mu+\sqrt{\tau_D/r}$, as predicted by Eq. \eqref{eq:Asympt_r}.

The behavior of $ \langle T_\mu(r;x)\rangle $ in other regimes can be explained from the one just analyzed. It is sufficient to consider the previous curve from some $ r_0>0 $. For example, the regime presented in Fig. \ref{fig:Line_cMFPT1}(b) can be understood from the previous one, by starting at $ r_m<r_0<r^*_\mu $, and the shape of the curve in Fig. \ref{fig:Line_cMFPT2}(c) has the same explanation as the previous case for $ r_0>r_M $. Of course, we are not claiming that we observe equivalent curves in this way, only that they their behaviors have the same qualitative explanation.

As a final remark, we emphasize that the behavior of the system depends on the size of the two time scales $ \tau_\mu=1/\mu $ and $ \tau_D=x^2/(4D) $. We have shown that, for fixed $ \tau_D $, different regimes are encountered by considering different values of $ \tau_\mu $. However, we could have obtained the same description by fixing $ \tau_\mu $ and varying $ \tau_D $. In other words, the transition to the different regimes presented above can also occur, for a given value of mortality, as the initial position $ x $ or the diffusion constant $ D $ varies.

\subsection{The conditional MFPT is not minimal at the rate which makes it equal to the standard deviation}
\begin{figure*}
	\subfloat{
		\includegraphics[width=.4\textwidth]{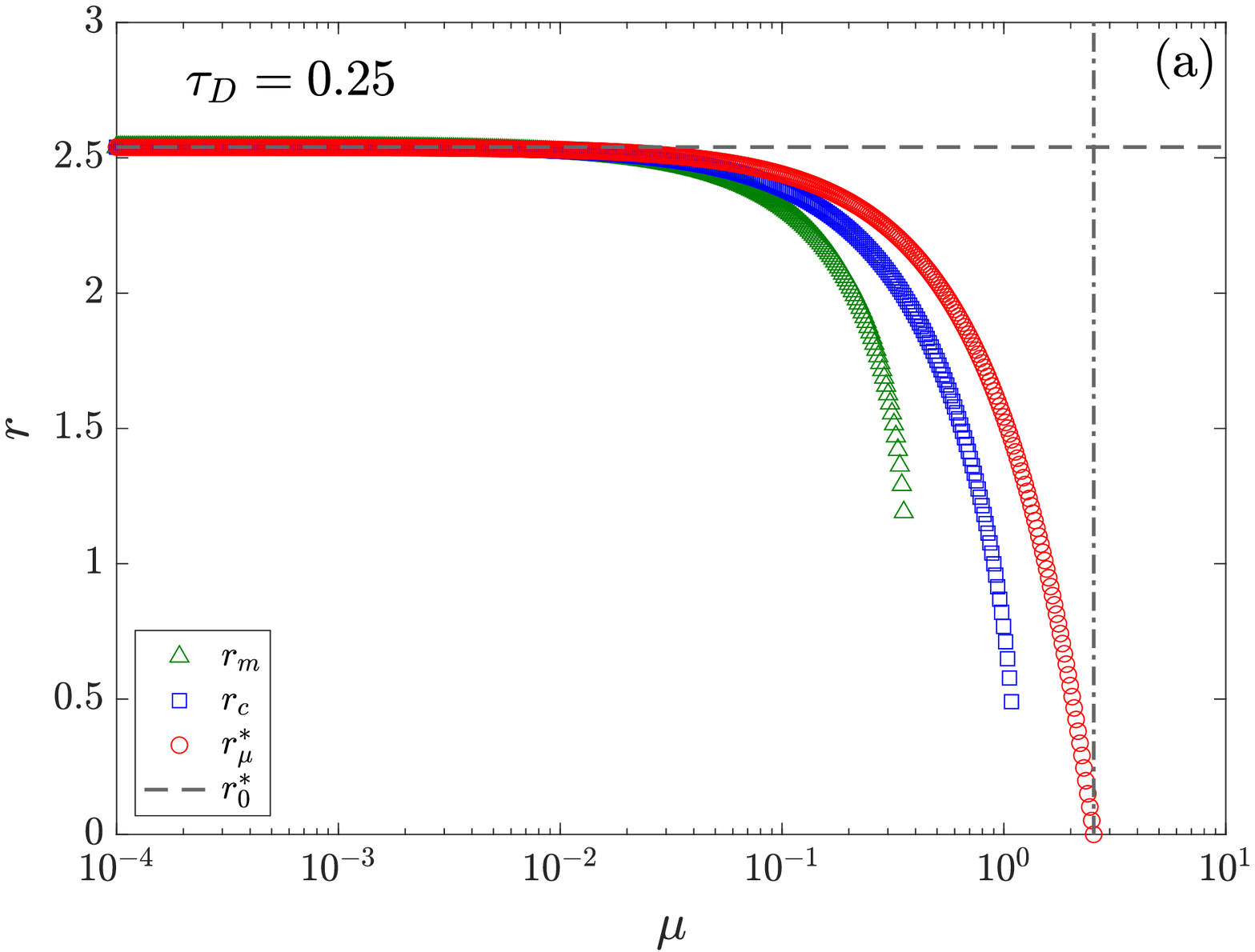}%
	}\quad
	\subfloat{
		\includegraphics[width=.4\textwidth]{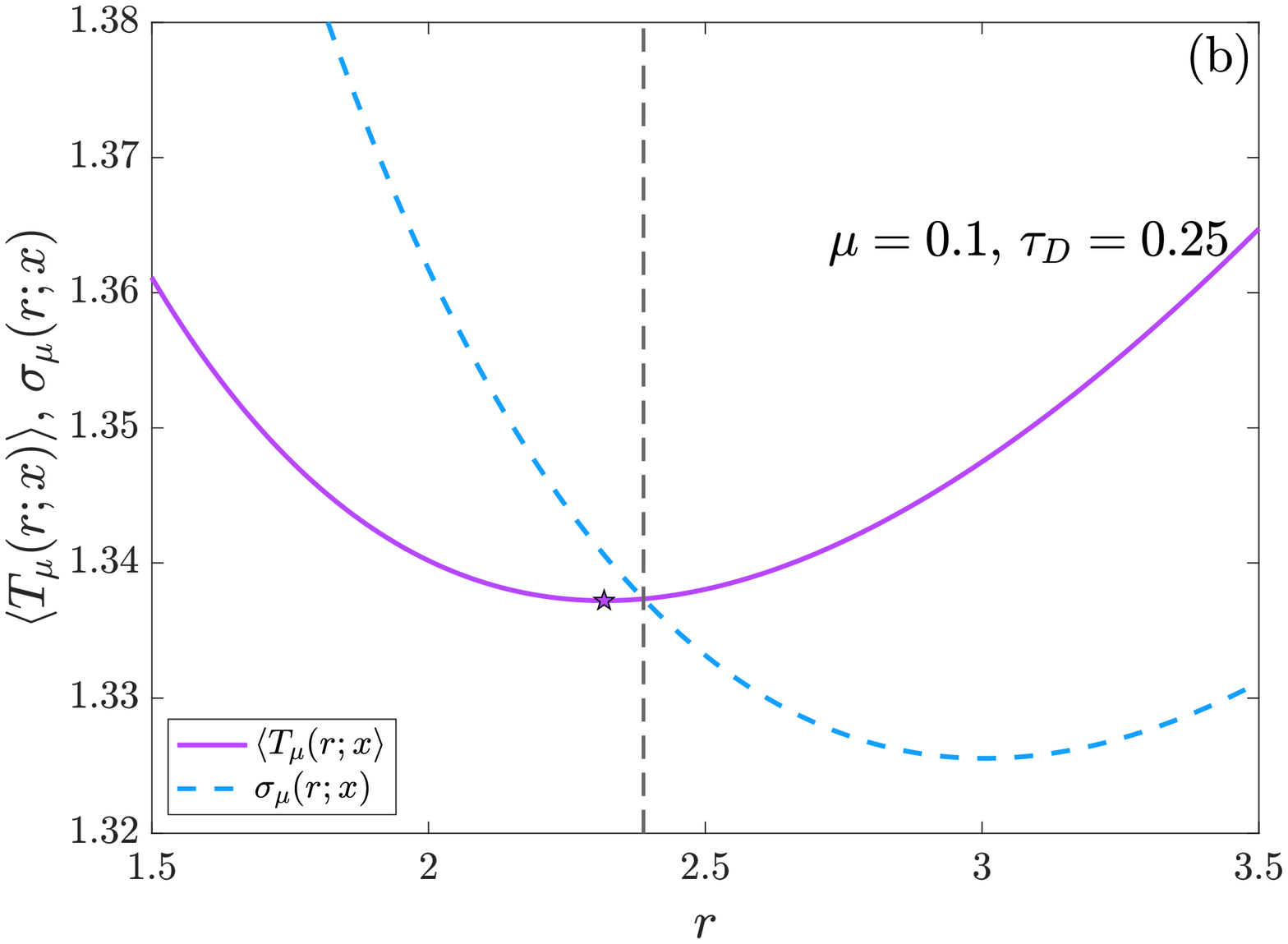}%
	}\hfill
	\caption{(a) Behavior of $ r_m $ (green triangles), $ r_c $ (blue squares) and $ r_\mu^* $ (red circles) versus $ \mu $. For each $ \mu $, we consistently find $ r_m<r_c<r_\mu^* $. For $ \mu\to0 $, all three converge to $ r_0^* $ (dashed horizontal line). Note that $ r_m $, $ r_c $ and $ r_\mu^* $ exist in different ranges of $ \mu $: For mortality rates $ \mu\gtrsim0.35 $ we can not find any $ r_m $, for $ \mu\gtrsim1.1 $ there is no $ r_c $, while $ r_\mu^* $ exists only for $ \mu\leq r_0^* $ (note the dashed-dotted vertical line at $ \mu=r_0^* $). (b) Conditional MFPT (purple solid line) and standard deviation (blue dashed line) for $ \mu=0.1 $. The position of the minimum of $ \langle T_\mu(r;x)\rangle $ (marker), $ r_m\approx 2.3172 $, clearly differs from the rate at which the two curves are equal, $ r_c\approx2.3884 $. Also note that $ \sigma_\mu(r_m;x)>\langle T_\mu(r_m;x)\rangle $.}
	\label{fig:Line_rs}
\end{figure*}

One of the most interesting results of stochastic resetting concerns a general property exhibited by the optimal resetting rate. It has been proven that for an arbitrary FPT process, subject to stochastic restarts with exponentially distributed time intervals, the relative fluctuation in the FPT, when the restart rate is optimal, is exactly unity \cite{Reu-2016}. In other words, if we introduce the coefficient of variation $ \CV(r)=\sigma_0(r)/\langle T_0(r)\rangle $, we must have
\begin{equation}\label{eq:Reu}
	\CV(r_0^*)=\frac{\sigma_0(r_0^*)}{\langle T_0(r_0^*)\rangle }=1.
\end{equation}
This relation has been well-established for different resetting systems \cite{RayMonReu-2019,RayReu-2020,RayReu-2021}. Deviations from this result may occur if the time intervals between resetting events are not exponentially distributed \cite{Reu-2016,PalReu-2017}, or due to the time cost to perform a single step in discrete-time random walk models \cite{BonPal-2021,RiaBoyHer-2022,RAD-2022-Gill}. However, as observed previously, we also find that this general result does not hold when we introduce mortality in the system. For instance, by considering the case $ \mu=0.1 $, we previously found that the rate $ r_c\approx2.3884 $ that makes the relative fluctuation equal to unity is slightly above $ r_m\approx2.3172 $. The same analysis can be repeated for different $ \mu $, and we consistently find $ r_m<r_c $. This happens both when the minimum is local and when it is global. Furthermore, we always have $ r_c<r_\mu^* $, hence
\begin{equation}\label{eq:Line_ineq_rates}
	r_m<r_c<r_\mu^*.
\end{equation}
Moreover, by looking at Fig. \ref{fig:Line_cMFPT1}(b), we see that there may exist $ r_c $ even in those cases where there is no minimum for $ \langle T_\mu(r;x)\rangle $.

A complete picture is presented in Fig. \ref{fig:Line_rs}(a), where we have evaluated numerically $ r_m $, $ r_c $ and $ r_\mu^* $ for different values of $ \mu $. In every case, we find that the chain of inequalities in Eq. \eqref{eq:Line_ineq_rates} is satisfied. As previously anticipated, $ r_m $, $ r_c $ and $ r_\mu^* $ exist in different domains, and in particular, there is a range of $ \mu $, which we estimate as $ 0.35\lesssim\mu\lesssim1.1 $, where the conditional MFPT has no minimum for positive $ r $, yet we can find $ r_c $ such that $ \sigma_{\mu}(r_c;x)=\langle T_\mu(r_c;x)\rangle $.

For $ \mu\to0 $, however, we should recover the general result known for immortal systems. Indeed, we see that all the numerical values converge to $ r_0^* $, which, we recall, is the rate that minimizes the MFPT. Thus, we can say that as long as the lifetime is infinite, an arbitrary FPT process subject to stochastic Poissonian resetting exhibits the property described by Eq. \eqref{eq:Reu}, viz., there is a unique rate $ r_0^* $ that minimizes the MFPT and makes the relative fluctuation in the FPT equal to unity. If we consider instead a finite lifetime, the probability of reaching the target, the relative fluctuation of the FPT and the conditional MFPT are optimized at different rates, whose values are $ \mu $-dependent.

To better understand the origin of the deviation from Eq. \eqref{eq:Reu}, let us consider the expression for $ \langle T_\mu(r;x)\rangle $ of Eq. \eqref{eq:cMFPT_vs_Mom}. One can verify that the derivative with respect to $ \mu $ and the derivative with respect to $ r $ satisfy the following relation:
\begin{equation}\label{key}
	\frac{\partial\langle T_\mu(r;x)\rangle}{\partial r}=\frac{\langle T_0^2(\mu+r;x)\rangle}{2(1+\mu\langle T_0(\mu+r;x)\rangle)^2}+\frac{\partial\langle T_\mu(r;x)\rangle}{\partial \mu}.
\end{equation}
Recalling Eq. \eqref{eq:cSig_vs_cMFPT}, the previous equation is equivalent to
\begin{equation}\label{eq:cMFPT_der_vs_Sig}
	\frac{\partial\langle T_\mu(r;x)\rangle}{\partial r}=\frac{\langle T_0^2(\mu+r;x)\rangle}{2(1+\mu\langle T_0(\mu+r;x)\rangle)^2}-\sigma_\mu^2(r;x),
\end{equation}
hence, the critical points $ r^* $ of $\langle T_\mu(r;x)\rangle$ satisfy
\begin{equation}
	\sigma_\mu^2(r^*;x)=\frac{\langle T_0^2(\mu+r^*;x)\rangle}{2(1+\mu\langle T_0(\mu+r^*;x)\rangle)^2}.
\end{equation}
Now, by using Eqs. \eqref{eq:cMFPT_vs_MFPT} and \eqref{eq:Reu_der}, we can translate this equation into a condition involving the coefficient of variation of the mortal system,
\begin{equation}\label{eq:Crit_Points}
	\CV_\mu(r^*)=\frac{\sqrt{\langle T_0(\mu+r^*;x)\rangle^2-\partial_\mu\langle T_0(\mu+r^*;x)\rangle}}{\langle T_0(\mu+r^*;x)\rangle+\mu\partial_\mu\langle T_0(\mu+r^*;x)\rangle},
\end{equation}
which, in general, is different from Eq. \eqref{eq:Reu}. Moreover, we also have a counterexample to a universal result, valid for any arbitrary restart protocol \cite{PalReu-2017}, affirming that at the optimal resetting rate
\begin{equation}\label{key}
	\CV(r_0^*)\leq 1.
\end{equation}
Instead, Fig. \ref{fig:Line_rs}(b) shows a case where at $ r=r_m $ the standard deviation is larger than the conditional MFPT. For the sake of completeness, we also point out that the previous inequality is recovered for $ r_\mu^* $. Once again, we can trace this back to the presence of mortality, that splits the optimal rate $ r_0^* $ for the immortal system in three different rates. Hence, the properties of the immortal system at $ r_0^* $ may diverge from those of the mortal system at $ r_m $.

The derivative given by Eq. \eqref{eq:cMFPT_der_vs_Sig} is displayed in Fig. \ref{fig:Line_cMFPT1}(d), for the case $ \mu=0.1 $. We used the expression for the variance of Eq. \eqref{eq:Line_sig}, while the second moment $ \langle T_0^2(p;x)\rangle $ can be easily obtained from Eq. \eqref{eq:Reu_der}. We recall that in this case the conditional MFPT is characterized by three critical points, see Fig. \ref{fig:Line_cMFPT1}(a). Indeed, we find three values of $ r $ for which the derivative vanishes: The first, which corresponds to a local maximum, is encountered at $ \overline{r}_M\approx0.0881 $; the second, which corresponds to the global minimum, is found at $ r_m\approx2.3172 $; the third is at $ r_M\approx279.1657 $, corresponding to the global maximum. This essentially confirms the condition for the critical points expressed by Eq. \eqref{eq:Crit_Points}, and explains why we observe deviations from the general result of Eq. \eqref{eq:Reu}.

\section{Summary and Conclusions}\label{s:Concl}
In this paper we have revisited the first-passage problem to the origin of one-dimensional diffusion undergoing stochastic resetting at constant rate $ r $, extending it to the case where the particle lifetime is a random time $ t $ distributed according to $ \psm(t)=\mu\exp(-\mu t) $, with $ \mu>0 $ defined as the mortality rate. This introduces the new time scale $ \tau_\mu=1/\mu $ into the problem, corresponding to the average lifetime of the particle, in addition to the two already present, namely, the average time interval between two resetting events, $ \tau_r=1/r $, and the diffusive time scale $ \tau_D=x^2/(4D) $. Depending on the relative magnitude of the various time scales, the system exhibits completely different properties that may diverge significantly from the behavior in absence of mortality.

The first quantity we investigated is the probability that the process ends by reaching the target before the particle decays, $ \mathcal{E}(\mu,r;x) $. Interestingly, this quantity is closely related to the MFPT of immortal particles, as described by Eq. \eqref{eq:E_vs_MFPT}. This allowed us to deduce that if the latter can be minimized for a certain resetting rate $ r_0^* $, then the former is maximized for $ r_\mu^*=r_0^*-\mu $, assuming that that $ \mu<r_0^* $. Thus, starting from the MFPT in absence of mortality, one can define the ranges of $ \mu $ within which resetting can increase the probability of a successful search.

We then considered the first two moments of the FPT, to provide a statistical description in terms of its mean and variance. These moments are obtained by averaging only over those processes that actually reach the target, so they are called conditional moments. Notably, the conditional MFPT, $ \langle T_\mu(r;x)\rangle $, can be derived from $ \mathcal{E}(\mu,r;x) $, see Eq. \eqref{eq:cMFPT_vs_E}. A first consequence due to mortality is that $ \langle T_\mu(r;x)\rangle $ has finite limits for both $ r\to0 $ and $ r\to\infty $, which makes the optimization problem much more complicated. In general, the behavior of the conditional MFPT is strongly affected by the value of the mortality rate, and we were able to identify four different regimes where $ \langle T_\mu(r;x)\rangle $ and the corresponding $ \mathcal{E}(\mu,r;x) $ have different properties.

The case in which mortality is so low as to allow both maximizing the probability and minimizing the conditional MFPT is certainly of great interest, and it is also the one that is most reminiscent of what is observed in absence of mortality. An example is the case $ \mu=0.1 $ that we examined earlier, displayed in Fig. \ref{fig:Line_cMFPT1}(a). Indeed, $ \langle T_\mu(r;x)\rangle $ has the global minimum for a certain resetting rate $ r_m $, which is preceded and followed by two maxima, a local one and the global one. The presence of these two maxima is due to mortality: In fact, in the presence of an infinite lifetime, two divergences would be observed instead. Note that $ r_m $ does not coincide with the rate that maximizes the probability, $ r_\mu^* $. Thus, one must decide whether it is preferable to increase the probability of a successful search, resulting in a slightly higher conditional MFPT, or to lower the MFPT as much as possible, while also decreasing the probability. Another effect due to mortality is the deviation from the result of Eq. \eqref{eq:Reu}. Indeed, we found that the relative fluctuation in the FPT is not unitary at $ r_m $; instead, this condition is achieved at a different rate $ r_c $, and we have $ r_m<r_c<r_\mu^* $.

The other cases are equally interesting in that they show how various mortality regimes can introduce significant differences from a description based on infinite lifetimes. Finally, let us stress that, except for the extreme mortality regime, the conditional MFPT always has a global maximum for some finite resetting rate $ r_M $. This may be relevant for some applications where, instead of minimizing the MFPT, one is instead interested in increasing it as much as possible, for example, when storing hazardous species whose leakage is to be reduced \cite{GreRup-2017}. Of course, it is intuitive that resetting at higher and higher rates increases more and more the MFPT. But the constraint of a finite average lifetime causes this growth to stop at a well-defined resetting rate $ r_M $. It is true that increasing the rate above $ r_M $ decreases the probability of leakage, but it also decreases the average time, and so, again, one must consider which of the two situations is more favorable.

We believe we have highlighted many interesting features peculiar to first-passage problems with resetting and mortality. Although we have focused only on one-dimensional diffusion, this paper can be a starting point for future work examining different and possibly more complex problems than the one analyzed here.

\appendix
\section{Derivative of the conditional MFPT}\label{a:Der}
In this Appendix we report some explicit calculations regarding the derivative with respect to the resetting rate of the conditional MFPT. We begin with the expression given by Eq. \eqref{eq:MFPT} in the main text: 
\begin{multline}\label{eq:a_MFPT}
	\langle T_\mu(r;x)\rangle=\frac{1+r\LT{f}'_0(x,\mu+r)}{\mu+r\LT{f}_0(x,\mu+r)}-\frac{1}{\mu+r}\\
	-\frac{\partial \ln\LT{f}_0(x,\mu+r)}{\partial r},		
\end{multline}
where
\begin{equation}\label{key}
	\LT{f}_0'(x,p)=\frac{\partial\LT{f}_0(x,p)}{\partial p}.
\end{equation}
The derivative with respect to $ r $ of Eq. \eqref{eq:a_MFPT} reads
\begin{multline}\label{eq:a_cMFPT_der}
	\frac{\partial\langle T_\mu(r;x)\rangle}{\partial r}=\frac{\LT{f}_0'(x,\mu+r)+r\LT{f}_0''(x,\mu+r)}{\mu+r\LT{f}_0(x,\mu+r)}\\
	-\frac{[\LT{f}_0(x,\mu+r)+r\LT{f}_0'(x,\mu+r)][1+r\LT{f}_0'(x,\mu+r)]}{[\mu+r\LT{f}_0(x,\mu+r)]^2}\\
	+\frac{1}{(\mu+r)^2}-\frac{\partial^2\ln\LT{f}_0(x,\mu+r)}{\partial r^2},
\end{multline}
which depends on $ \LT{f}_0(x,p) $, $ \LT{f}_0'(x,p) $ and $ \LT{f}_0''(x,p) $. We can also obtain an alternative expression involving $ \langle T_0(r;x)\rangle $ and its derivatives by considering Eq. \eqref{eq:cMFPT_vs_MFPT} of the main text, which yields
\begin{multline}\label{eq:a_cMFPT_der_vs_MFPT}
	\frac{\partial\langle T_\mu(r;x)\rangle}{\partial r}=\frac{1}{1+\mu\langle T_0(\mu+r;x)\rangle}\Bigg[\mu\frac{\partial^2\langle T_0(\mu+r;x)\rangle}{\partial r^2}\\
	+\frac{1-\mu^2\partial_r\langle T_0(\mu+r;x)\rangle}{1+\mu\langle T_0(\mu+r;x)\rangle}\frac{\partial\langle T_0(\mu+r;x)\rangle}{\partial r}\Bigg],
\end{multline}
where we used the fact that
\begin{equation}\label{key}
	\frac{\partial\langle T_0(\mu+r;x)\rangle}{\partial\mu}=\frac{\partial\langle T_0(\mu+r;x)\rangle}{\partial r}.
\end{equation}
Note that, by setting the lhs of Eq. \eqref{eq:a_cMFPT_der_vs_MFPT} to zero, one obtains the condition
\begin{multline}\label{eq:a_cMFPT_st_points}
	\frac{1-\mu^2\partial_r\langle T_0(\mu+r;x)\rangle}{1+\mu\langle T_0(\mu+r;x)\rangle}\frac{\partial\langle T_0(\mu+r;x)\rangle}{\partial r}\\
	=-\mu\frac{\partial^2\langle T_0(\mu+r;x)\rangle}{\partial r^2},
\end{multline}
which is a general of equation for the critical points of $ \langle T_\mu(r;x)\rangle $. From this equation, we see that if $ \langle T_0(r;x)\rangle $ has a local minimum at $ r^*>\mu $, with $ \partial^2\langle T_0(r;x)\rangle/\partial r^2>0 $ at $ r=r^* $, then for $ r=r^*-\mu$ the rhs of Eq. \eqref{eq:a_cMFPT_der_vs_MFPT} is positive and thus the conditional MFPT is locally increasing. Conversely, if $ \langle T_0(r;x)\rangle $ has a local maximum at $ r^*>\mu $, with $ \partial^2\langle T_0(r;x)\rangle/\partial r^2<0 $ at $ r=r^* $, then $ \langle T_{\mu}(r;x)\rangle $ is locally decreasing. In particular, if there exists $ r_0^*>\mu $ that minimizes $ \langle T_0(r;x)\rangle $, then $ r^*_\mu=r_0^*-\mu $ maximizes the probability of a successful search, see Sec. \ref{s:E}, but increases the conditional MFPT relative to lower resetting rates.

The sign of the derivative $ \partial\langle T_\mu(r;x)\rangle/\partial r $ for $ r\to0 $ indicates whether the introduction of resetting with infinitesimally small rate increases or decreases the conditional MFPT. By evaluating Eq. \eqref{eq:a_cMFPT_der} at $ r=0 $ we get
\begin{multline}\label{key}
	\left.\frac{\partial\langle T_\mu(r;x)\rangle}{\partial r}\right|_{r=0}=\frac{\LT{f}_0(x,\mu)}{\mu}\left[\frac{\LT{f}_0'(x,\mu)}{\LT{f}_0(x,\mu)}-\frac1\mu\right]+\frac1{\mu^2}\\
	-\frac{\partial^2\ln\LT{f}_0(x,\mu)}{\partial \mu^2},
\end{multline}
which is equivalent to
\begin{multline}\label{key}
	\left.\frac{\partial\langle T_\mu(r;x)\rangle}{\partial r}\right|_{r=0}=-\frac{\mathcal{E}(\mu,0;x)}{\mu}\left[\langle T_\mu(0;x)\rangle+\frac1\mu\right]\\
	+\frac{1}{\mu^2}-\sigma^2_\mu(0;x).
\end{multline}
By introducing the coefficient of variation $ \CV_\mu $ of the reset-free mortal process
\begin{equation}\label{key}
	\CV_\mu=\frac{\sqrt{\langle T^2_\mu(0;x)\rangle-\langle T_\mu(0;x)\rangle^2}}{\langle T_\mu(0;x)\rangle}=\frac{\sigma_{\mu}(0;x)}{\langle T_\mu(0;x)\rangle},
\end{equation}
and the dimensionless quantity $ \zeta_\mu$
\begin{equation}\label{key}
	\zeta_\mu=\frac{\langle T_\mu(\infty;x)\rangle}{\langle T_\mu(0;x)\rangle},
\end{equation}
the condition of having a negative slope, so that the introduction of resetting decreases the MFPT, reads
\begin{equation}\label{eq:aCond_neg_slope}
	\CV^2_\mu>\left[1-\mathcal{E}(\mu,0;x)\right]\zeta_\mu^2-\mathcal{E}(\mu,0;x)\zeta_\mu.
\end{equation}

\end{document}